\documentclass[usenatbib,fleqn]{mnras}

\usepackage{mathptmx}

\usepackage{amsmath}

\usepackage{graphicx}

\usepackage{hyperref}

\usepackage{xspace}

\usepackage{color}

\usepackage{upgreek}

\newcommand{\epow}[1]{\mathrm{e}^{#1}}

\renewcommand{\pi}{\uppi}

\newcommand{\Mpc}{\,h^{-1}\mathrm{Mpc}}

\newcommand{\iMpc}{\,h\mathrm{Mpc}^{-1}}

\newcommand{\Msun}{\,h^{-1}\mathrm{M_\odot}}

\newcommand{\Om}{\Omega_\mathrm{m}}

\newcommand{\Ob}{\Omega_\mathrm{b}}
\newcommand{\Ow}{\Omega_w}

\newcommand{\dc}{\delta_\mathrm{c}}

\newcommand{\Dv}{\Delta_\mathrm{v}}

\newcommand{\eg}{e.g.,\xspace}
\newcommand{\ie}{i.e.\xspace}

\newcommand{\nbody}{$N$-body\xspace}
\newcommand{\LCDM}{$\Lambda$CDM\xspace}
\newcommand{\halofit}{\textsc{halofit}\xspace}

\newcommand{\emu}{\textsc{cosmic emu}\xspace}
\newcommand{\gadget}{\textsc{gadget-2}\xspace}

\newcommand{\camb}{\textsc{camb}\xspace}

\newcommand{\meadaddress}{\href{https://github.com/alexander-mead/collapse}{https://github.com/alexander-mead/collapse}\xspace}

\newcommand{\new}[1]{{\color{black}{#1}}}

\def\mnras{MNRAS}
\def\apj{ApJ}
\def\apjl{ApJ Let.}
\def\prd{Phys. Rev. D}
\def\aap{A\&A}
\def\jcap{J. Cosmol. Astropart. Phys.}
\def\apjs{ApJS}

\def\m@th{\mathsurround=0pt }
\def\eqalign#1{\null\,\vcenter{\openup1\jot\m@th\ialign{\strut\hfil$\displaystyle{##}$&$\displaystyle{{}##}$\hfil\crcr#1\crcr}}\,}
\def\gtsim{\mathrel{\lower0.6ex\hbox{$\buildrel {\textstyle >}\over {\scriptstyle \sim}$}}}
\def\ltsim{\mathrel{\lower0.6ex\hbox{$\buildrel {\textstyle <}\over {\scriptstyle \sim}$}}}

\title[Spherical collapse and formation hysteresis]{Spherical collapse, formation hysteresis and the deeply non-linear cosmological power spectrum}
\author[A. J. Mead]{A. J. Mead\thanks{E-mail: alexander.j.mead@googlemail.com} \\
Department of Physics and Astronomy, University of British Columbia, 6224 Agricultural Road, Vancouver, \new{BC V6T 1Z1}, Canada\\
Canadian Institute for Theoretical Astrophysics, University of Toronto, \new{ON M5S 3H8}, Canada\\
}

\date{Accepted 2016 September 9. Received 2016 September 8; in original form 2016 June 16}
\pagerange{\pageref{firstpage}--\pageref{lastpage}} 
\pubyear{2016}

\begin{document}
\maketitle

\label{firstpage}

\begin{abstract}
I examine differences in non-linear structure formation between cosmological models that share a $z=0$ linear power spectrum in both shape and amplitude, but that differ via their growth history. \nbody simulations of these models display an approximately identical large-scale-structure skeleton, but reveal deeply non-linear differences in the demographics and properties of haloes. I investigate to what extent the spherical-collapse model can help in understanding these differences, in both real and redshift space. I discuss how this is difficult to do if one attempts to identify haloes directly, because in that case one is subject to the vagaries of halo finding algorithms. However, I demonstrate that the halo model of structure formation provides an accurate non-linear response in the power spectrum, but only if results from spherical collapse that include formation hysteresis are properly incorporated. I comment on how this fact can be used to provide per cent level accurate matter power spectrum predictions for dark energy models for $k\leq5\iMpc$ by using the halo model as a correction to accurate \LCDM simulations. In the appendix I provide some fitting functions for the linear-collapse threshold ($\dc$) and virialized overdensity ($\Dv$) that are valid for a wide range of dark energy models. I also make my spherical-collapse code available at \meadaddress.
\end{abstract}


\section{Introduction}
\label{sec:introduction}

The collapse of an isolated, spherical top hat under the action of gravity is a well studied problem in large-scale structure cosmology \citep{Gunn1972}, and a rare example of non-linear gravitational evolution that can be solved exactly. In an Einstein-de Sitter (flat, $\Om=1$) cosmological model, it is well known that this `spherical model' predicts that non-linear collapse should have occurred when linear theory would predict a perturbation to have reached the `critical overdensity' value $\dc\simeq1.686$. This number is the height of the barrier that must be crossed in analytical calculations of the halo mass function (\eg \citealt{Press1974}, \citealt{Bond1991}, \citealt*{Sheth2001}). If one additionally assumes that the top hat will virialize then the resulting halo is predicted to be $\Dv\simeq 178$ times denser than the background. This virial criterion is used to inform halo finding algorithms in cosmological \nbody simulations, where it is standard to define haloes as objects that are $\sim 200$ times denser than the background.

The spherical-collapse model can be solved when $\Om\neq 1$, and deviations from the Einsten-de Sitter results are found. These solutions may provide insights into how gravitational collapse changes in non-standard cosmologies (\eg dark energy: \citealt{Mota2004}, \citealt{Percival2005}, \citealt{Pettorino2008}, \citealt{Wintergerst2010}, \citealt{Pace2010}, \citealt{Tarrant2012}; modified gravity: \citealt{Schmidt2009a,Schmidt2010a}, \citealt{Brax2010}, \citealt{Li2012b}, \citealt{Barreira2013}; massive neutrinos: \new{\citealt{Ichiki2012}}, \citealt{LoVerde2014}). The spherical-collapse calculation was first carried out for non-flat cosmologies by \cite{Lacey1993} and for \LCDM by \cite*{Eke1996} who show that $\dc$ decreases a small amount (by less than one per cent) as the universe evolves to $\Om(a)<1$, while $\Dv$\footnote{I always define $\Dv$ with respect to the background matter density, \emph{not} the critical density.} increases more drastically (approximately doubles). The spherical-collapse model is clearly very simplistic, and it is not obvious that accurate spherical-model calculations have a role in accurate predictions for cosmological observables such as the halo mass function or $n$-point statistics. Should we treat the spherical model as nothing other than a provider of a general trend, or should we continue to carry out accurate spherical collapse calculations for new cosmological models? Answering these questions by investigating haloes in \nbody simulations is complicated because halo finding algorithms often bunch particles into haloes in a way that is inconsistent with the spherical model. For example, haloes may be defined with a user-set spherical-overdensity (SO) criterion, independent of the background cosmology, or via a friends-of-friends (FoF) algorithm with a fixed linking length. It is not obvious that either algorithm provides the `correct' definition of a halo, given that SO assumes sphericity where haloes are obviously not spherical, while FoF can link visually distinct structures via a bridge and has non-trivial dependences on mass resolution and halo profile \citep{Warren2006,More2011}. These issues contribute to conflicting claims regarding universality; to what extent can the mass function be expressed as a redshift- and cosmology-independent function of the variance in the linear density field? Some authors have demonstrated the mass function to be a close-to-universal function, but only when haloes are identified with a fixed SO threshold \citep[\eg][]{Tinker2008} or fixed FoF linking length \citep[\eg][]{Warren2006,Watson2013}. There are conflicting claims in the literature regarding universality when different identification criteria are used. \cite{Jenkins2001}, \cite{Tinker2008} and \cite{Courtin2011} report a breakdown of universality, but more recently claims to the contrary have been made by \cite{Despali2016} who see mass function universality across a range of \LCDM cosmological parameters when using $\Dv$ predictions to define haloes with a SO criterion.

In this paper I test the spherical-collapse model in a cosmological setting by examining differences in \nbody simulations that have a fixed $z=0$ linear power-spectrum shape and amplitude, but that have different growth histories. The fixing of linear modes isolates non-linear differences and I question whether the spherical model is useful for understanding these residuals. The most obvious tests involve investigating haloes directly, as has been done previously \citep[\eg][]{Courtin2011}, but in doing so one encounters problems associated with the vagaries of halo definition. I circumvent this by comparing the matter power spectra measured from the simulations to predictions from the halo model of structure formation when I vary the halo-model ingredients. The question that I wish to address is; does the cosmology dependence of $\dc$ and $\Dv$ have any real relevance for real structure formation as the underlying cosmology is changed?

This paper is ordered as follows: in Section~\ref{sec:spherical_collapse}, I review the spherical-collapse model. In Section~\ref{sec:halo_identification}, I discuss halo identification methods and why the spherical-collapse model is difficult to test if one is tied to a particular halo definition. Simulations of cosmological models with different forms of dark energy, but fixed $z=0$ linear power spectra, are then presented in Section~\ref{sec:simulations} and their non-linear differences are discussed. In Section~\ref{sec:halo_model}, I discuss the halo model, which is my method of choice for comparing spherical-collapse predictions to simulations. Results of these comparisons are then presented in Section~\ref{sec:results}, which is followed by a summary in Section~\ref{sec:summary}.

\section{The spherical-collapse model}
\label{sec:spherical_collapse}

It is possible to follow the evolution of a spherically symmetric `top-hat' perturbation in an otherwise featureless universe using both linear and non-linear theory. Clearly the linear treatment will not be accurate once the perturbation is sufficiently developed, but successful approaches to understanding the mass function \citep{Press1974,Bond1991} use the idea of haloes forming once the \emph{linear} density field passes a threshold value, and have proved to be extremely useful in describing features of the non-linear Universe.

I will be interested in the matter overdensity, $\delta$, defined relative to the background matter density via $1+\delta=\rho/\bar\rho$. Under the assumption that a spherical hat remains a spherical hat throughout its evolution the non-linear equation of motion for the overdensity is:
\begin{equation}
\ddot\delta+2H\dot\delta-\frac{4}{3}\frac{\dot\delta^2}{1+\delta}=\frac{3}{2}\Om(a)H^2\delta(1+\delta)\ ,
\label{eq:spherical_tophat}
\end{equation}
where a dot denotes a time derivative and $H$ is the Hubble parameter. If $\Om=1$ and $\delta\ll 1$, then $\delta\propto a$, but growth accelerates as the perturbation develops. If the perturbation is sufficiently extreme it will reach a maximum size and then collapse\footnote{Not all top hats collapse. For example, if $\Lambda$ comes to dominate at late times a perturbation can be prevented from collapsing by the compensating outward acceleration.}, defined as the time when $\delta\rightarrow\infty$, which occurs as a result of the exact spherical symmetry. On linearizing equation~(\ref{eq:spherical_tophat}) the standard expression for the linear evolution of an arbitrary perturbation configuration is recovered:
\begin{equation}
\ddot g+2H\dot g=\frac{3}{2}\Om(a)H^2 g\ .
\label{eq:linear_tophat}
\end{equation}
where $g$ is the linearized $\delta$. The linear collapse threshold, $\dc$, is defined as the value that the linear field has reached when $\delta\rightarrow\infty$ in the non-linear calculation. In an Einstein-de Sitter cosmology $\dc\simeq 1.686$, and the trend with cosmological parameters is that $\dc$ decreases very slightly as $\Om(a)$ decreases. For example, $\dc\simeq1.676$ for $\Om(a)=0.3$ in \LCDM, a decrease of $\simeq0.6$ per cent.

The virialized density contrast of the collapsed hat, $\Dv$, is calculated by applying the virial theorem at the time of collapse, which sets the radius of the structure. Anisotropy in any realistic perturbation will prevent it from collapsing to a singularity, and violent relaxation \citep{Lynden-Bell1967} will then allow virialization. In an Einstein-de Sitter cosmology the radius of the collapsed halo is half the radius of the perturbation at turnaround, giving $\Dv\simeq 178$.  $\Dv$ changes in different cosmologies, with the trend that it increases as $\Om(a)$ decreases in \LCDM (I always define $\Dv$ relative to the background matter density). Virialization is a more tricky concept with a cosmological constant \citep{Lahav1991}  because this provides an outward radial force that counters self-gravity in a halo to some extent, meaning the halo should collapse to a smaller size to virialize. For other dark energy models, a physical model is required before virialization can be fully described \citep{Mota2004,Maor2005}. Even if dark energy is taken to be exactly homogeneous, the amount of dark energy in a collapsed halo changes over time in dynamical models, meaning that virialization may never be achieved \citep{Percival2005}. For realistic haloes (\ie, not top hats) the importance of the outward force provided by dark energy will be profile dependent, and it is not obvious that a solution to virialization in the top-hat case will be relevant to real structure formation. Therefore, in this paper I ignore the direct effect of dark energy on virialization and simply set the virial radius of haloes to be half the maximum perturbation radius, independently of the underlying cosmological parameters. For \LCDM this gives $\Dv\simeq 310$ for $\Om=0.3$, an increase of $\simeq75$ per cent compared to $\Om=1$. If instead one includes the cosmological constant in the virilization process \citep[\eg][]{Lahav1991,Eke1996,Bryan1998} then $\Dv\simeq 337$, and so the difference between these approaches is $\simeq10$ per cent.

\section{Halo identification}
\label{sec:halo_identification}

The spherical-collapse model makes predictions that relate to collapsed haloes, and to compare directly with \nbody simulations haloes need to be identified from the particle distribution. Typically, these are found using either a FoF (\citealt{Davis1985}) or a SO (\citealt{Press1974}) algorithm. Different halo-finding techniques are compared in \cite{Knebe2011}. In FoF, a linking length is set and haloes are defined as sets of particles that are within the linking length of at least one other particle in the set. This will lead to non-spherical structures being identified as haloes, and objects that look visually like two distinct haloes may be joined by a bridge. If haloes follow an isothermal profile, $\rho\propto r^{-2}$, a linking-length in terms of the mean inter-particle separation of $b=0.2$ translates into an enclosed overdensity of $\Dv\simeq 180$, which is close to the Einstein-de Sitter predictions from the spherical model, and so $b=0.2$ is quite standard. One might be tempted to think that fixing $b$ translates into a surface density criterion for haloes identified from particles; however, \cite{Warren2006} showed via experiments on Poisson realizations of isothermal haloes that internal overdensities of FoF haloes range from $200$ to $400$ in a way that depends on the number of particles resolving the halo, and that no clear iso-density surface is picked out. Therefore the FoF mass assigned to a halo will be a function of the mass resolution and \cite{Warren2006} provide a resolution-dependent correction for this effect. \cite{More2011} conduct similar experiments with the more realistic \citeauthor*{Navarro1997} (NFW; \citeyear{Navarro1997}) haloes and find the average enclosed overdensity of FoF haloes is also a function of the halo density profile, so the FoF mass will also be profile dependent. In contrast, SO algorithms pick a halo centre (via either the centre-of-mass of an FoF group, or the minimum in the density field or gravitational potential in a region, or using some iterative mix of these) and then `grow' a sphere until this encloses a user-specified overdensity. The exact details of an algorithm mean that some particles may contribute to more than one halo and that aspherical particle distributions get classified as spheres, when in reality they may have a very different density structure and be strongly aspherical. SO algorithms may also fail to assign some dense regions to any halo if the region falls between two parent haloes. In either FoF or SO gravitationally unbound particles may also be removed from a halo.

One semi-obvious fact that should be borne in mind is that isolated spherical haloes with obvious boundaries do not exist in reality. From simulations one observes that an initially Gaussian distribution of linear fluctuations grows and collapses into a `cosmic web' with sharp peaks in the density populating a cosmic skeleton. These peaks attract material and merge with one another, and the density profile around a peak changes over time. A halo only comes into existence when a somewhat arbitrary boundary is drawn around these peaks; sometimes these boundaries will encompass multiple peaks (\eg halo substructure or FoF overlinking) and sometimes a spherical shape will be forced upon an obviously non-spherical density distribution (\eg SO finders). It should also be remembered that different halo identification methods can be suitable for different tasks. For example, haloes defined using an SO algorithm with $\Dv=500$ may relate to peaks in the density field that can be identified by X-ray observations, but might not be the same haloes that are useful for understanding massive cluster abundance or the details of the small-scale matter distribution.

Independently of the details of halo identification, the mass distribution function of haloes has been shown to be expressible as an approximately universal function in terms of $\nu=\dc/\sigma(M)$, where $\sigma(M)$ is the variance in the linear density field when smoothed on a comoving scale that encloses a mass $M$. Successful analytical calculations of the mass function take the linear density field smoothed on successively smaller scales, and equate the tail of the distribution of overdensity beyond the critical $\dc$ with the cumulative mass function for haloes greater than that filter mass \citep{Press1974,Bond1991}. This approach works well in explaining the broad form of the mass function, but fails in detail. The response has been the development of various fitting functions for the mass function, based on the \cite{Press1974} idea, but calibrated to high resolution \nbody simulations \citep[\eg][]{Sheth1999,Jenkins2001,Warren2006,Reed2007,Tinker2008,Bhattacharya2011,Watson2013,Despali2016}. The details of the mass function therefore depend on the halo identification scheme. Often these mass functions are parameterized in terms of $\sigma(R)$, rather than $\nu$, and if $\dc$ is incorporated its cosmology dependence is often neglected. Even though the changes in $\dc$ with cosmology are small ($\simeq$ 1 per cent) they can have large effects because of the exponential form of the mass function. For example, for a mass function $\propto \mathrm{e}^{-\nu^2/2}$, a one per cent decrease in $\dc$ leads to a $\simeq5$, $10$ and $20$ per cent increase in the abundance of $\nu=2$, $3$ and $4$, objects respectively.

\cite{Sheth1999} produced the first calibrated, universal mass function by fitting to haloes identified in simulations; the authors include the cosmology dependence of $\dc$ and identify haloes using SO with a cosmology-dependent overdensity criterion (Sheth \& Tormen, private communication), but error bars are large because simulations were of low resolution compared to those available today. \cite{Jenkins2001} identified haloes using FoF and report a universal mass function, but only if a fixed linking length is set independently of cosmology, and their mass function is parameterized as a function of $\sigma(M)$ only (\ie $\dc$ is phased out). \cite{Warren2006} correct FoF masses to account for the finite number of halo particles, but do not investigate the cosmology dependence of the mass function, and parameterize in terms of $\sigma$ rather than $\nu$. \cite{Reed2007} use FoF with $b=0.2$ and report non-universality in the mass function; they account for this by including the effective spectral index of the power spectrum at the collapse scale in their fitting function, making it non-universal. \cite{Bhattacharya2011} also note non-universality with $b=0.2$ FoF haloes over a range of $w$CDM cosmologies and show that this is not alleviated by including $\dc$ in the mass function. \cite{Tinker2008} use SO to define haloes with a variety of fixed $\Dv$ and also report non-universality, their mass function is parameterized as in terms of $\sigma(M)$ only and includes a universality-breaking redshift dependence. However, the authors also note that the mass function appears closer to universal if they use FoF haloes -- a similar conclusion is reached by \cite{Watson2013}. Most recently, \cite{Despali2016} use SO with the spherical-model dependence in $\Dv$ to define haloes, include $\dc$ changes in their fitting functions, and report universality in the mass function across a range of \LCDM cosmologies. 


Attempting to isolate the effect of $\dc$ on the mass function is made difficult because different cosmological models often have different linear spectrum amplitudes and shapes, which changes $\sigma(M)$, and therefore the effect of a $\dc$ change alone is difficult to isolate. Despite this, \cite{Courtin2011} demonstrated that, for a fixed FoF halo definition, universality \emph{is} enhanced in dark energy models if one takes the cosmological dependence of $\dc$ into account in the mass function, but only for the most massive haloes, while deviations from universality remain at lower masses and are not remedied. The authors also note that the remaining non-universality is correlated with the spherical-collapse $\Dv$ prediction, but are unable to generate a universal FoF mass function using the spherical model to inform the $\Dv$ choice. However, it is certainly possible that enhanced universality might be obtained if more correct cosmology-dependent halo definitions were used \citep{Despali2016}, although due to the observations of \cite{Warren2006} and \cite{More2011} this is hard to implement in practice with FoF. 

Attempting to directly compare spherical-model $\Dv$ predictions to simulations is hard because one must first define a halo before asking questions about its overdensity. SO algorithms set the $\Dv$ threshold manually, whereas with FoF one is subjected to the vagaries of the FoF method that particularly affect halo size. Therefore, even if one did measure the internal densities of FoF haloes it would be unclear how to interpret the results.



\section{The simulations}
\label{sec:simulations}

\begin{table}
\begin{center}
\caption{The cosmological parameters of the simulations used in this paper. Dynamical dark energy is parameterized via $w(a)=w_0+(1-a)w_a$ and is taken to be spatially homogeneous, thus only affecting the background expansion. All simulations use $512^3$ particles in cubes of size $L=200\Mpc$, and start from initial conditions with identical mode phases but amplitudes adjusted to ensure $\sigma_8=0.8$ at $z=0$. The shape of the linear spectrum used to generate the initial conditions is identical in each case, and was generated using \camb \protect\citep{Lewis2000} with cosmological parameters $\Om=0.3$, $\Omega_w=1-\Om$, $\Ob=0.05$, $h=0.7$, $n_\mathrm{s}=0.96$ and $w=-1$. For each cosmological model I ran three different realizations of the initial conditions. I also show the spherical-model parameters $\dc$ and $\Dv$ from a numerical calculation for each cosmological model at $a=1$.}
\begin{tabular}{c c c c c c c}
\hline
Cosmology & $\Om$ & $\Omega_w$ & $w_0$ & $w_a$ & $\dc$ & $\Dv$ \\
\hline
\LCDM & 0.3 & 0.7 & $-1$    & $0$     &  $1.6755$ & $310.1$\\
EdS     & 1.0 & 0.0 & --         & --          & $1.6865$ & $177.4$\\
Open   & 0.3 & 0.0 & --         & --          & $1.6513$ & $402.0$\\
DE1     & 0.3 & 0.7 & $-0.7$ & $0$     & $1.6695$ & $342.7$\\
DE2     & 0.3 & 0.7 & $-1.3$ & $0$     & $1.6787$ & $282.4$\\
DE3     & 0.3 & 0.7 & $-1$    & $0.5$  & $1.6724$ & $318.5$\\
DE4     & 0.3 & 0.7 & $-1$    & $-0.5$ & $1.6773$ & $301.6$\\
DE5     & 0.3 & 0.7 & $-0.7$ & $-1.5$  & $1.6774$ & $313.3$\\
DE6     & 0.3 & 0.7 & $-1.3$ & $0.5$ & $1.6771$ & $290.1$\\
\hline
\end{tabular}
\label{tab:sims}
\end{center}
\end{table}

\begin{figure}
\begin{center}
\includegraphics[angle=270,width=8.5cm]{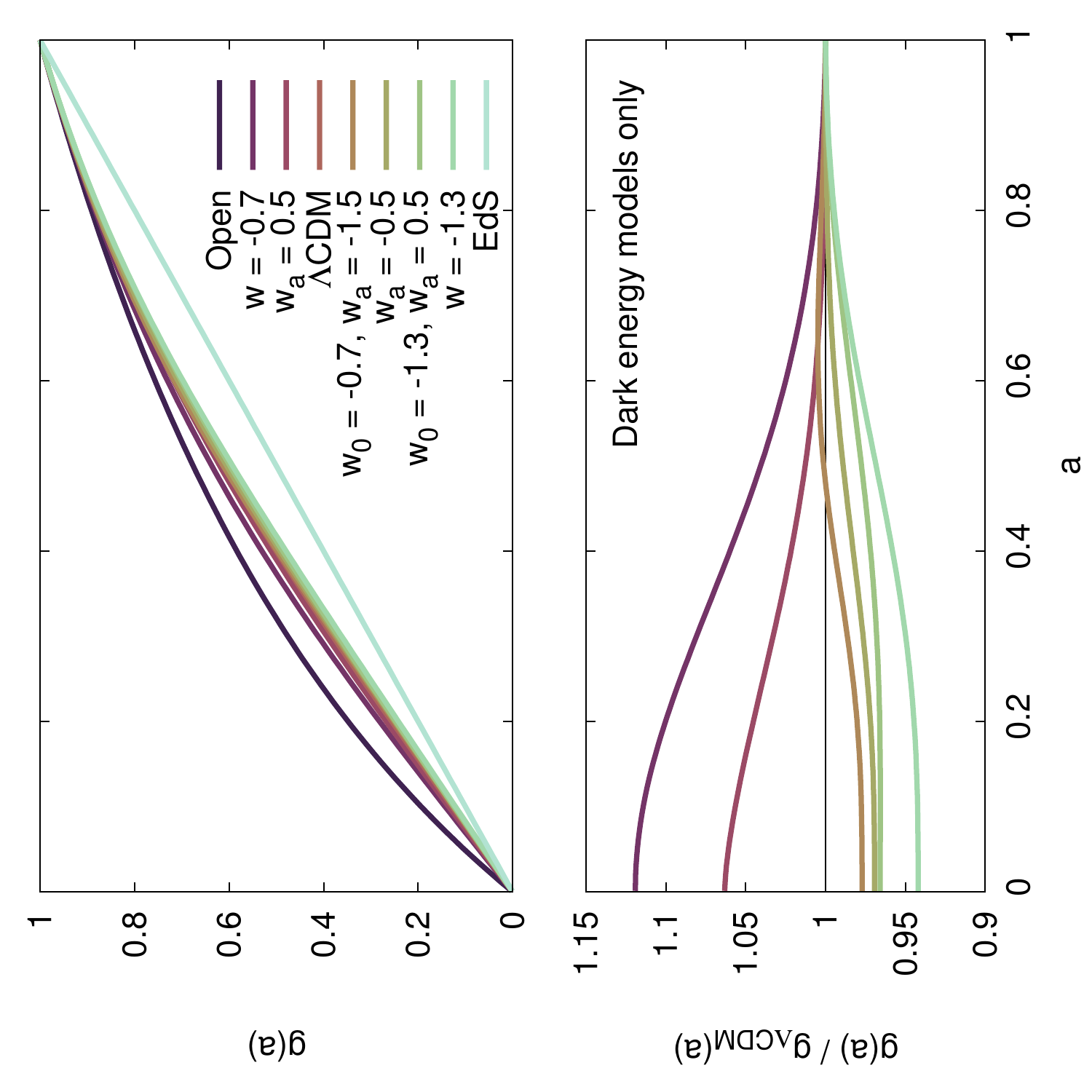}
\end{center}
\caption{The linear-theory growth factor versus the scale factor for the cosmological models considered in this work. The top panel shows the growth factors normalized so that $g(a=1)=1$ while the lower panel shows the ratio of growth factors for the dark energy cosmology with respect to \LCDM. The cosmological models investigated in this paper share a linear theory amplitude at $a=1$, but all have different growth histories. Therefore any differences in non-linear structure formation must be due to hysteresis.}
\label{fig:growth}
\end{figure}

To gauge the utility of the spherical-collapse model it is convenient to design simulations with differences confined to predictions for spherical model parameters as much as possible. I do this using the simulations with cosmological parameters listed in Table~\ref{tab:sims}, which have different values of $\Om$, $\Ow$ and different underlying spatially homogeneous dark energy\footnote{Although for $w\neq-1$ fluid dark energy \emph{must} have perturbations to be mathematically consistent \citep[\eg][]{Weller2003} if the sound speed is high (\eg simple scalar field models, where the sound speed is the speed of light) then dark-energy perturbations are erased below the horizon, which is huge compared to the scales considered in this paper.}, parameterized via $w(a)=w_0+(1-a)w_a$. Simulation initial conditions are set with the same linear power spectrum shape (taken from \camb; \citealt{Lewis2000}), with the initial amplitudes adjusted such that the simulations will have identical linear mode amplitudes at $z=0$. Initial condition phases are identical, so that comparisons may be made in the absence of cosmic variance. For each cosmological model I ran three different realizations using the \gadget code \citep{Springel2005b} in cubes of side length $L=200\iMpc$, which is a good compromise for getting high enough resolution at small scales without sacrificing too much large-scale power. Although not directly relevant to this work, the constant $w$ models DE1 and DE2 and time-varying model DE3 are well outside the $2\sigma$ confidence region from \cite{Planck2015XIV}, but the other dark-energy models lie approximately within the ellipse. The degeneracy line of this confidence region is defined by $w(a)$ models that share an angular distance to the last-scattering surface. Unfortunately, these models also have a similar linear growth history because both the growth and distance to last scattering depend on $H(a)$, which is very similar for these models. This makes them particularly difficult to distinguish observationally because both the CMB and large-scale structure will appear very similar.

The linear growth functions for the cosmologies in Table~\ref{tab:sims} are shown in Fig.~\ref{fig:growth}, where it can be seen that all have different growth histories. In reality, they could be distinguished from linear information alone, using redshift evolution, but here I confine my investigations to $a=1$, when their linear content is identical.

\begin{figure*}
\begin{center}
\includegraphics[angle=270,width=17.5cm]{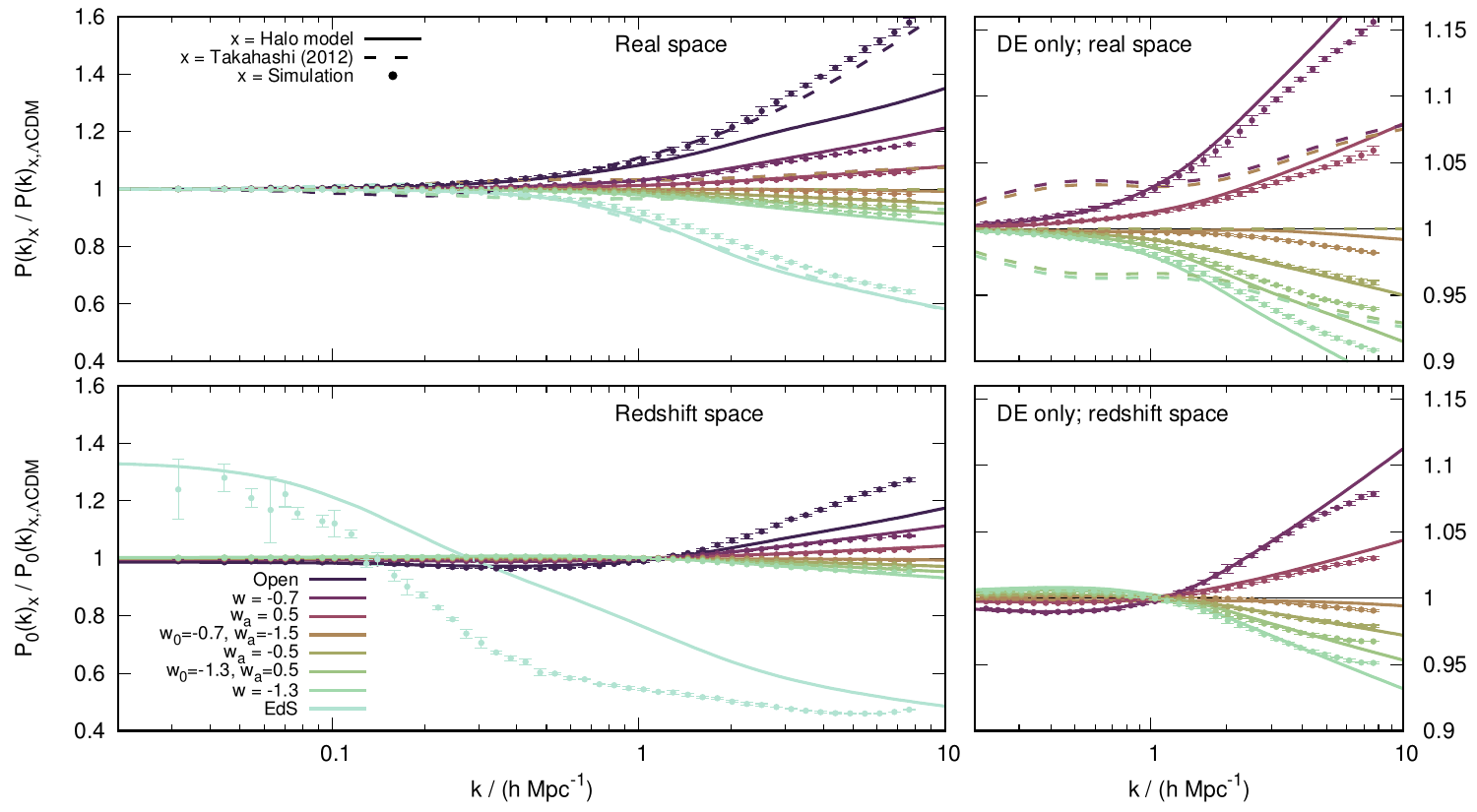}
\end{center}
\caption{The power-spectrum response (ratio to \LCDM) measured in simulations is shown as the points with error bars (error-on-the-mean from three realizations) in both real (top panels) and redshift space (bottom panels). The left-hand panels show the full range of scales probed by the simulations whereas the right-hand panels show only the dark energy models and are zoomed-in on the smallest scales. The solid lines show the main result of this paper, which is the halo-model prediction for the response when the choice of ingredients has been made carefully and the spherical-collapse values for $\dc$ and $\Dv$ are included, this response is particularly accurate for the dark energy models as can be seen in the right-hand panels. In the left-hand panels it can be seen that the general trend of the halo-model response is matched to the simulation in each case, even for the extreme open and EdS cosmologies, which have very different growth histories compared to \LCDM. The ordering of the halo-model curves is correct in all cases. The mild $\dc$ cosmology dependence means that the halo mass distribution that go into the halo model will be slightly different for each cosmology, whereas the $\Dv$ and $c(M)$ dependence ensure that halo profiles will differ. The halo model has not been fit to the data in any way. The dashed curve in the upper panels shows the response from \halofit of \protect\cite{Takahashi2012}, which works well for the open and Einstein-de Sitter cosmologies, but performs poorly for the dark energy cosmologies, with the \halofit response being exactly unity for any cosmologies that share a $w(a=1)$ value.}
\label{fig:power}
\end{figure*}

Throughout this paper I will only be interested in the matter power spectrum `response'; the ratio of power from one method to the power for a fiducial cosmological model from that \emph{same} method. \ie, a simulation measurement divided by another simulation measurement for the fiducial cosmological model, or a halo-model prediction divided by the halo-model prediction for the \emph{same} fiducial cosmological model. The response says nothing about the absolute accuracy of a prediction scheme, but instead demonstrates how effectively the scheme responds to changes in cosmological parameters. The power-spectrum response measured from these simulations at $a=1$ is shown in Fig.~\ref{fig:power} as the points with error bars (error-on-the-mean from three realizations). The spectra were computed using FFTs on a $1024^3$ mesh and have not been shot noise corrected because this is negligible for the scales shown. I show results for both real space and the monopole of redshift space. The small size of the error bar at high-$k$ demonstrates that this response is quite insensitive to the random seed used for the initial conditions. I also checked that the response was insensitive to resolution by running some $100$ and $400\Mpc$ boxes, getting very similar results for the response. In real space one can see that power agrees for each cosmology at large scales, as expected, but there are deviations that begin around $k=0.2\iMpc$ leading to quite different power spectra for $k>1\iMpc$. At large scales in redshift space there is a large difference for the Einstein-de Sitter cosmology, which arises because that model has a very different \cite{Kaiser1987} boost; there is both a large scatter because there are few modes and a low bias because some large-scale velocity structure is missing from these simulations due to the small $L=200\Mpc$ box. The large-scale response better matches the expectation from the difference in Kaiser factors if a larger simulation volume is used. For the other models for $k>0.2\iMpc$ there are non-linear differences that initially follow the Kaiser factor, but then swap sense to larger differences for $k>1\iMpc$, which must arise from different halo structure and different strength of the Fingers-of-God, but the power-spectrum differences are suppressed somewhat compared to real space.

A similar set of dark energy simulations have been investigated by \cite*{McDonald2006}, who look at differences in structure formation between cosmologies with different, constant $w$ values compared to an equivalent \LCDM model. These authors noted similar differences at non-linear scales to those seen in Fig.~\ref{fig:power} (the results presented here are entirely consistent with their results) and also note that the power-spectrum response is much more accurately provided by simulations than the raw value of the power -- a result obtained by detailed convergence testing. They suggest that this accuracy arises because some types of numerical error cancel in the ratio. These authors also provide a fitting function for a $w$-dependent correction to the power spectrum, and note that this correction is quite insensitive to the linear spectrum shape.

If one assumes that the linear power spectrum uniquely determines the non-linear, then all of these cosmologies should have equal non-linear power despite their different growth histories -- Fig.~\ref{fig:power} shows that this assumption is obviously incorrect. The original \halofit of \cite{Smith2003} would predict that all of the dark energy cosmologies share a non-linear spectrum, but predicts differences between the \LCDM, open and Einstein-de Sitter cosmologies through explicit dependence on curvature and matter density. Predictions from the updated \halofit of \cite{Takahashi2012} are shown as the dashed curves in Fig.~\ref{fig:power}, where it can be seen to match the response for the open and Einstein-de Sitter cosmologies well, but the prediction for the dark energy cosmologies is comparatively poor. \halofit of \cite{Takahashi2012} contains explicit $w$ dependence, but therefore predicts all cosmologies that share a $w$ value to have identical non-linear spectra if $w$ in the \citealt{Takahashi2012} fitting function is interpreted as $w(a)$, which produces the most accurate \halofit results according to \citealt{Mead2016}. The augmented halo model of \cite{Mead2015b,Mead2016} \emph{does} predict differences between these cosmologies because the concentration--mass relation in the halo model depends on the growth history. However, for the cosmologies with fixed linear theory investigated in this paper the \cite{Mead2015b} model performs only marginally better than \halofit, and is not shown. I now turn the attention of the reader to a predictive scheme that can be used to understand the origin of the differing responses. A preview of this result show via the solid lines in Fig.~\ref{fig:power}, which match the simulations particularly well for the dark energy cosmologies.

\section{The halo-model power spectrum}
\label{sec:halo_model}

The power spectrum of statistically isotropic density fluctuations depends only on $k=|\mathbf{k}|$ and is given by
\begin{equation}
P(k)=\langle|\delta_\mathbf{k}|^2\rangle\ ,
\label{eq:P(k)}
\end{equation}
where the average is taken over all modes with the same modulus. I find it convenient to define the dimensionless quantity $\Delta^2$:
\begin{equation}
\Delta^2(k)=4\pi L^3\left(\frac{k}{2\pi}\right)^3 P(k)\ ,
\label{eq:Delta2_definition}
\end{equation}
which gives the fractional contribution to the variance per logarithmic interval in $k$ in a cube of volume $L^3$. If the overdensity field is filtered on a comoving scale $R$, the variance is
\begin{equation}
\sigma^2(R,z)=\int_0^{\infty}\Delta^2(k,z)\, T^2(kR)\;\mathrm{d}\ln{k}\ ,
\label{eq:variance}
\end{equation}
with the window function
\begin{equation}
T(x)=\frac{3}{x^3}(\sin{x}-x\cos{x})\ ,
\label{eq:top_hat}
\end{equation} 
corresponding to smoothing with a spherical top-hat. A mass scale may be related via $M=4\pi R^3\bar\rho_\mathrm{m} /3$, where $\bar\rho_\mathrm{m}$ is the average matter density.

\begin{figure*}
\begin{center}
\includegraphics[width=17.5cm]{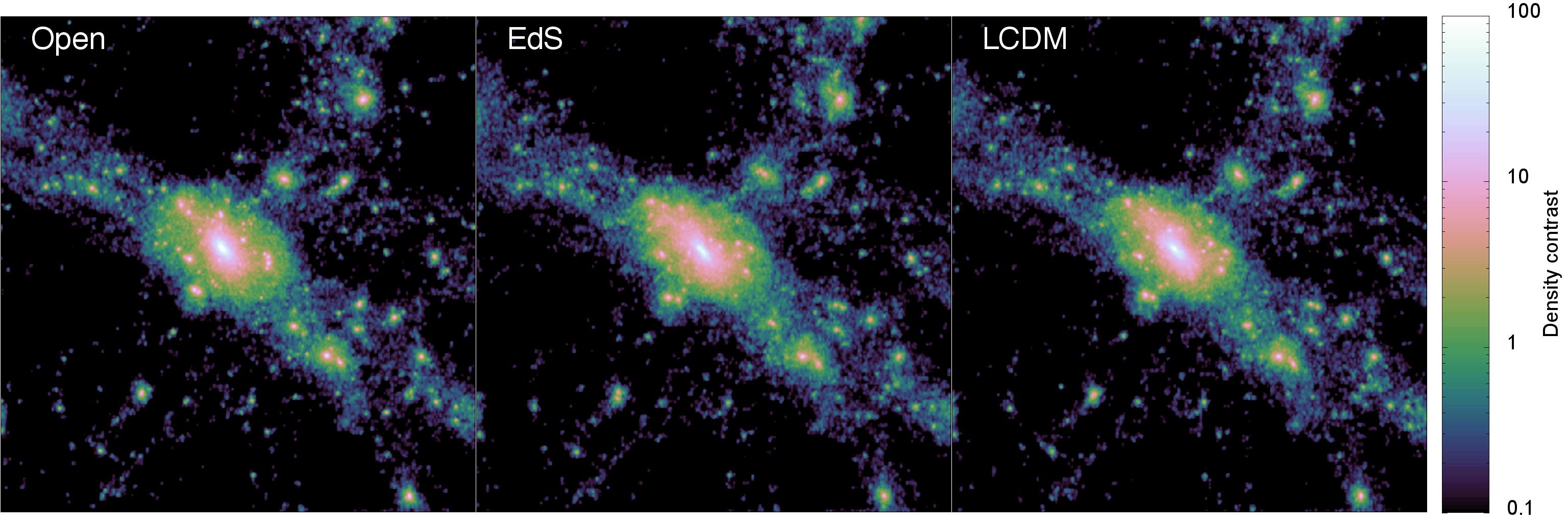}
\end{center}
\caption{The projected 2D density field around the same density peak in three simulations. The central halo is $\sim 10^{15}\Msun$, but the exact mass assigned to the object, and the number of objects found, would depend on the halo-finding algorithm. Each panel shows the projected density along one face of the same $15\Mpc$ cube. In each simulation the peak exists within the same large-scale structure skeleton, a consequence of the identical linear power spectra and matched simulation initial conditions. What is striking is that, despite the different growth histories, the haloes themselves appear very similar at $a=1$, and that even sub-haloes and infall can be matched between the different panels. Measured differences in mass function may therefore reflect subtle differences in halo boundary and profile, rather than major differences in the properties of the peaks that define halo centres. These subtly different profiles must also give rise to the different power spectra seen in Fig.~\ref{fig:power}.}
\label{fig:density}
\end{figure*}

It has been demonstrated that halo density structure is well approximated by the profiles of NFW:
\begin{equation}
\rho(r)=\frac{\rho_\mathrm{N}}{(r/r_\mathrm{s})(1+r/r_\mathrm{s})^2}\ ,
\label{eq:nfw}
\end{equation}
where $r_\mathrm{s}$ is a scale radius that roughly separates the core of the halo from the outer portion, $\rho_\mathrm{N}$ is a normalization, and the halo is truncated at the virial radius: $M=4\pi r_\mathrm{v}^3\Dv\bar\rho_\mathrm{m} /3$. $r_\mathrm{s}$ is usually expressed via the halo concentration parameter $c=r_\mathrm{v}/r_\mathrm{s}$. I use the concentration--mass relations of \cite{Bullock2001}:
\begin{equation}
c(M,z)=A\frac{1+z_\mathrm{f}(M)}{1+z}\times\frac{g(z\to\infty)}{g_{\Lambda\mathrm{CDM}}(z\to\infty)}\ , 
\label{eq:bullock_cm}
\end{equation} 
where the ratio of linear growth function\footnote{Here $g(z)$ is normalized such that $g(z=0)=1$} to that of \LCDM is a correction advised by \cite{Dolag2004} for dark-energy cosmologies. The calculation of the halo formation redshift, $z_\mathrm{f}$, as a function of mass, is described in \cite{Bullock2001}, and crucially depends on the formation history of the halo, so there is some hysteresis whereby haloes retain a memory of their formation time. It has been demonstrated \citep[\eg][]{Kazantzidis2004} that the internal velocity structure of haloes is well described by a mass-dependent Gaussian distribution (independent of radius), with a dispersion given by the virial theorem
\begin{equation}
\sigma^2_\mathrm{v}=\frac{GM}{2r_\mathrm{v}}\ .
\end{equation}

The halo model may be used to generate a prediction for the matter power spectrum in both real \citep[][]{Peacock2000,Seljak2000,Cooray2002}  and redshift space \citep{White2001}. The power is decomposed into one- and two-halo terms that separate clustering arising from structure within individual haloes from that arising from between different haloes:
\begin{equation}
\Delta^2(k,\mu)=\Delta_{2H}^2(k,\mu)+\Delta_{1H}^2(k,\mu)\ ,
\label{eq:full_power}
\end{equation}
where $\mu=\cos\theta$ and $\theta$ is the angle of the mode to the line of sight.

The simplest form of the power-spectrum calculation makes the assumptions that haloes are smooth and spherical, with a profile that is uniquely determined by the mass. A two-halo term can be written down by additionally assuming that haloes of mass $M$ are linearly biased with respect to the linear density field, $\delta_\mathrm{H}=b(M)\delta$:
\begin{equation}
\Delta_{2H}^2(k,\mu)=(1+\new{f_\mathrm{g}}\mu^2)^2\Delta_\mathrm{lin}^2(k)\left[\int_0^\infty f(\nu)b(\nu)W(M,k,\mu)\;\mathrm{d}\nu\right]^2\ ,
\label{eq:2halo}
\end{equation}
where the pre-factor is the redshift-space factor of \cite{Kaiser1987}, \new{$f_\mathrm{g}=\mathrm{d}\ln{g}/\mathrm{d}\ln{a}$, $g$ is the linear growth function}, $\nu=\dc/\sigma(M)$ and $f(\nu)$ is the mass function. $W(M,k,\mu)$ is the normalized spherical Fourier transform of the halo density profile, convolved in real space by the Gaussian velocity distribution term
\begin{equation}
W(M,k,\nu)=\mathrm{e}^{-k^2\mu^2\sigma_\mathrm{v}^2/2}\frac{1}{M}\int_0^{r_\mathrm{v}}\frac{\sin(kr)}{kr}\ 4\pi r^2\rho(r,M)\;\mathrm{d}r\ . 
\label{eq:halo_window}
\end{equation}
For the mass function I use the fitted function of \cite{Sheth1999}:
\begin{equation}
f(\nu)=A\left[1+\frac{1}{(a\nu^{2})^p}\right]\epow{-a\nu^2/2}\ ,
\label{eq:STmassfunction}
\end{equation}
with parameters $a=0.707$, $p=0.3$ and the normalization, $A$, constrained such that the integral of $f(\nu)$ over all $\nu$ must equal one: $A\simeq 0.2162$. Note that in fitting their mass function \cite{Sheth1999} use a cosmology-dependent $\Dv$ for SO halo identification, and also include the spherical-collapse $\dc$ dependence in the conversion between $M$ and $\nu$. I use the halo bias appropriate for the mass function assuming the peak-background split approach of \cite{Sheth1999}, which ensures that the integral in equation~(\ref{eq:2halo}) tends to unity as $k\rightarrow 0$. 

The one-halo term accounts for power that arises from inside haloes, and has the form of shot noise moderated by the density profile of the haloes:
\begin{equation}
\Delta_\mathrm{1H}^2(k,\mu)=4\pi\left(\frac{k}{2\pi}\right)^3\frac{1}{\bar\rho_\mathrm{m}}\int_0^\infty M f(\nu) W^2(M,k,\mu)\;\mathrm{d}\nu\ .
\label{eq:1halo}
\end{equation}

Note that if $\mu=0$ the spectrum is undistorted and equations~(\ref{eq:2halo}), (\ref{eq:halo_window}) and (\ref{eq:1halo}) reduce to the standard real-space expressions. In \cite{White2001} only the redshift-space monopole is considered, and the required average over $\mu$ is performed `under the integral' in equations~(\ref{eq:2halo}) and (\ref{eq:1halo}), but I calculate $\Delta(k,\mu)$ in full from the halo model and then compute multipoles afterwards. Note also that the integral correction to the linear power in equation~(\ref{eq:2halo}) that damps off power at small scales is relatively unimportant in real space \citep{Cooray2002}, but is \emph{much} more important in redshift space where the one-halo term is more subdued due to the fingers-of-God effect. I performed a simple check of the redshift-space calculation by comparing the ratio of real to redshift-space monopole power to the same ratio measured from simulations and found excellent agreement, very similar to that presented in \cite{White2001}.

The halo model is known \emph{not} to provide accurate matter power spectra when compared to \nbody simulations \citep[\eg][]{Smith2003,Valageas2011,Mead2015b} and deviations at the tens of per cent level are seen at non-linear scales. I have checked that similar discrepancies are also present in redshift space. This suggests that some features of the non-linear density field are missed by the simple form of the halo-model calculation (equations~\ref{eq:2halo} and \ref{eq:1halo}). These certainly include: unvirialized structures, non-linear (and stochastic) halo biasing, halo asphericity and tidal alignment, halo substructure and variation of halo properties at fixed mass. The main result of this paper is that, despite this lack of absolute accuracy, the response to changes in cosmological parameters that the halo model does predict \emph{is} accurate if one incorporates the spherical-collapse model, but only if one restricts oneself to cosmologies that share a linear theory.

\section{Results}
\label{sec:results}

The projected density field around the same density peak in three different simulations is shown in Fig.~\ref{fig:density}. The most striking feature of these plots is how similar the density fields look, even down to the details of the substructure, and the infalling objects. The most obvious difference is visible between the open and Einstein-de Sitter cosmologies, which are the two with the most different growth histories. One can see that the peaks in the open case are more defined, and the central cluster in the Einstein-de Sitter case is more diffuse, a consequence of it having formed more recently in time. I do not show the density field for the dark-energy simulations, because they all look so similar to that in \LCDM. Given that these simulations share a large-scale structure skeleton, the differences in power shown in Fig.~\ref{fig:power} must be due to these subtle differences in the halo properties, and this can be accounted for using the halo model.

\begin{figure*}
\begin{center}
\includegraphics[angle=270,width=18.4cm]{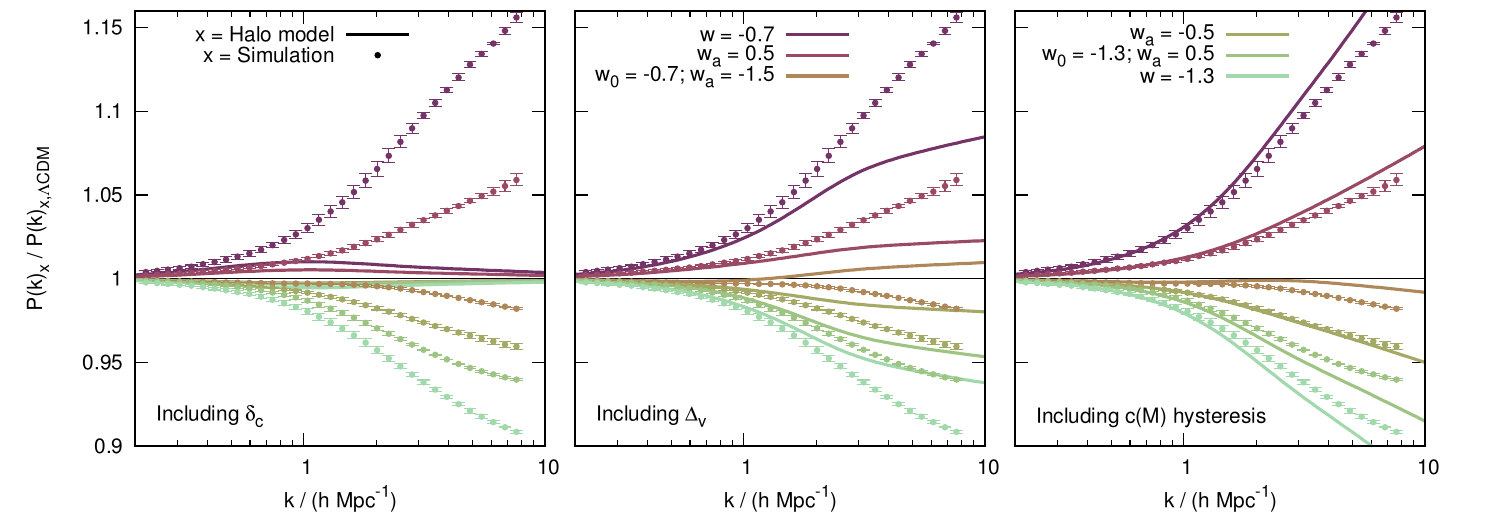}
\end{center}
\caption{The build-up of the halo-model prediction for the power-spectrum response (ratio of power to \LCDM) for the dark energy cosmologies. The points show the response as measured in the simulations (error-on-the-mean from three realizations), whereas the solid lines show the halo-model prediction. If the values $\Dv=200$ and $\dc=1.686$ are fixed, and the concentration--mass relation is a function of halo mass and redshift only, then halo-model predictions are identical and the response will be unity because all cosmologies share a linear theory. In the left-hand panel the spherical-model predictions for $\dc$ has been incorporated, leading to modest improvements. In the central panel the $\Dv$ predictions have been added, which leads to a larger improvement. The final result, including hysteresis in the concentration--mass relation, is shown in the right-hand panel, which is a carbon copy of the top-right panel of Fig.~\ref{fig:power}, where it can be seen that the halo-model response is accurate at the per cent level for $k<5\iMpc$.}
\label{fig:progression}
\end{figure*}

The simple halo-model prediction for the non-linear power spectrum (equations \ref{eq:full_power}, \ref{eq:2halo} and \ref{eq:1halo}) of a cosmological model will clearly depend on what ingredients are used. For example, if a mass function is used that depends on $\sigma(M)$ only (\ie omitting $\dc$), and $\Dv=200$ is fixed \citep[\eg][]{Jenkins2001,Warren2006,Tinker2008}, together with concentration--mass relations that are only functions of $M$ \citep[\eg][]{Neto2007,Duffy2008,Prada2012}, then the halo model will predict \emph{exactly} the same non-linear power for each cosmology, in obvious disagreement with Fig.~\ref{fig:power}. It is therefore important to use ingredients that can respond to the change in cosmology in the correct manner, such as those discussed in Section~\ref{sec:halo_model}. This change power-spectrum response as the halo-model ingredients are modified is shown as the solid lines in Fig.~\ref{fig:progression}. In the left-hand panel $\Dv=200$ is fixed and the concentration--mass relation is a function of $M$ and $z$ only, but the spherical-model predictions for $\dc$ have been included, which changes the relationship between $\nu$ and $M$ and leads to improvements in the predicted response (that would otherwise be unity for all models). In the central panel the spherical-collapse $\Dv$ predictions have been added, which changes halo profiles and leads to a larger improvement. The final result is shown in the right-hand panel (a carbon copy of the top-right panel of Fig.~\ref{fig:power}), where I use $\dc$ and $\Dv$ values from the spherical model and incorporate a concentration--mass relation that accounts for the different growth history (\citealt{Bullock2001}, with the \citealt{Dolag2004} correction). Although the halo model in its raw form is a poor match to simulations, the response allows us to `divide out' this inaccuracy. One can see this result in full in Fig.~\ref{fig:power} in real and redshift space, where almost perfect results are obtained for dark energy models when the predicted spherical-collapse values are used; changes in $\dc$ are important around $k<1\iMpc$, changes in $\Dv$ are important for $k\sim1\iMpc$ and cosmology dependence in the concentration--mass relation for $k>1\iMpc$. This figure demonstrates that spherical-model predictions \emph{are} relevant for the non-linear power spectrum, and that incorporating them within the halo model allows for an almost perfect understanding of non-linearity in the power spectrum for $k<7\iMpc$, at least for the dark-energy cosmologies. The results for the open cosmology are less impressive, but clearly the halo-model trend is in the correct direction, the comparative lack of precision may be because the concentration--mass relation is augmented to be accurate for dark energy cosmologies, rather than curved ones. The fact that the Einstein-de Sitter cosmology is well matched is probably because most cosmologies investigated tend to Einstein-de Sitter in the past, and many halo-model ingredients are fitted over a range of redshifts. The resulting power spectra responses shown in Fig.~\ref{fig:power} are accurate to one per cent for the dark energy cosmologies for $k\leq5\iMpc$, and this has involved no fitting parameters whatsoever. The redshift-space power spectrum of dark matter provides a neat second test for method, and is shown in the lower panel in Fig.~\ref{fig:power}, where one can see that the combination of Kaiser factor at large scales and halo-velocity dispersion at small scales again results in good halo-model predictions for the response, actually better than in real space for the dark energy models. The redshift-space Einstein-de Sitter response is that worst predicted by the halo model, which may be because the growth rate is very different, and the quasi-linear regime may depend non-trivially on cosmological parameters in a way that is not captured by the simple halo-model calculation. I note that these results are insensitive to the choice of mass function, and in fact the residual is slightly more accurate if I use the \cite{Press1974} function, which may be because that mass function has a closer relation with the \emph{spherical}\new{-collapse} model (note the \citealt{Sheth1999} form can be partly justified using the \emph{ellipsoidal}-collapse model of \citealt*{Sheth2001}).

The interpretation of these results is as follows: cosmological models that share a linear power spectrum will not generally share a non-linear spectrum, but they do have roughly equal `quasi-linear' ($k<0.3\iMpc$) spectrum. \new{This has been noted by many authors, including \cite{Zheng2002} and \cite{McDonald2006}. \cite{Nusser1998} show that the equations of motion for \nbody particles can be caste in a form that is almost independent of cosmological parameters if the linear growth rate is used as a time variable, which probably explains the quasi-linear accuracy.} The magnitude of residual differences at deeply non-linear scales is strongly correlated with the difference in growth history, which has been pointed out previously, including by\cite*{Francis2007}, \cite{Ma2007} and \cite{Alimi2010}. What is interesting is that the halo model, when combined with the spherical-collapse model, essentially accounts for residual differences in the non-linear power spectra, which must be because the spherical model successfully captures the hysteresis. This fact should be borne in mind when attempting to derive prediction schemes for the non-linear power, be they perturbation theory, effective field theory or halo model based.

\new{One might worry that the accuracy of the results for the response shown in Fig.~\ref{fig:progression} depend on the baseline \LCDM cosmological parameters. However, note that \cite{McDonald2006} showed that for constant $w$ models the response was only very weakly dependent on the \LCDM parameters and also note that the halo-model prediction for the response \emph{will} change as baseline cosmological parameters are varied because many of the halo-model ingredients change with cosmology. To double-check the results presented in this paper are robust to baseline changes I ran some additional simulations to measure the response with different \LCDM parameters. In these cases the power-spectrum response from the simulations was only very slightly different to that shown in Fig.~\ref{fig:progression} and the halo model prediction for the response was also very slightly different, however the halo-model response prediction was always accurate at a similar level. For brevity these extra results are not presented here.}

\begin{figure}
\begin{center}
\includegraphics[angle=270,width=8.5cm]{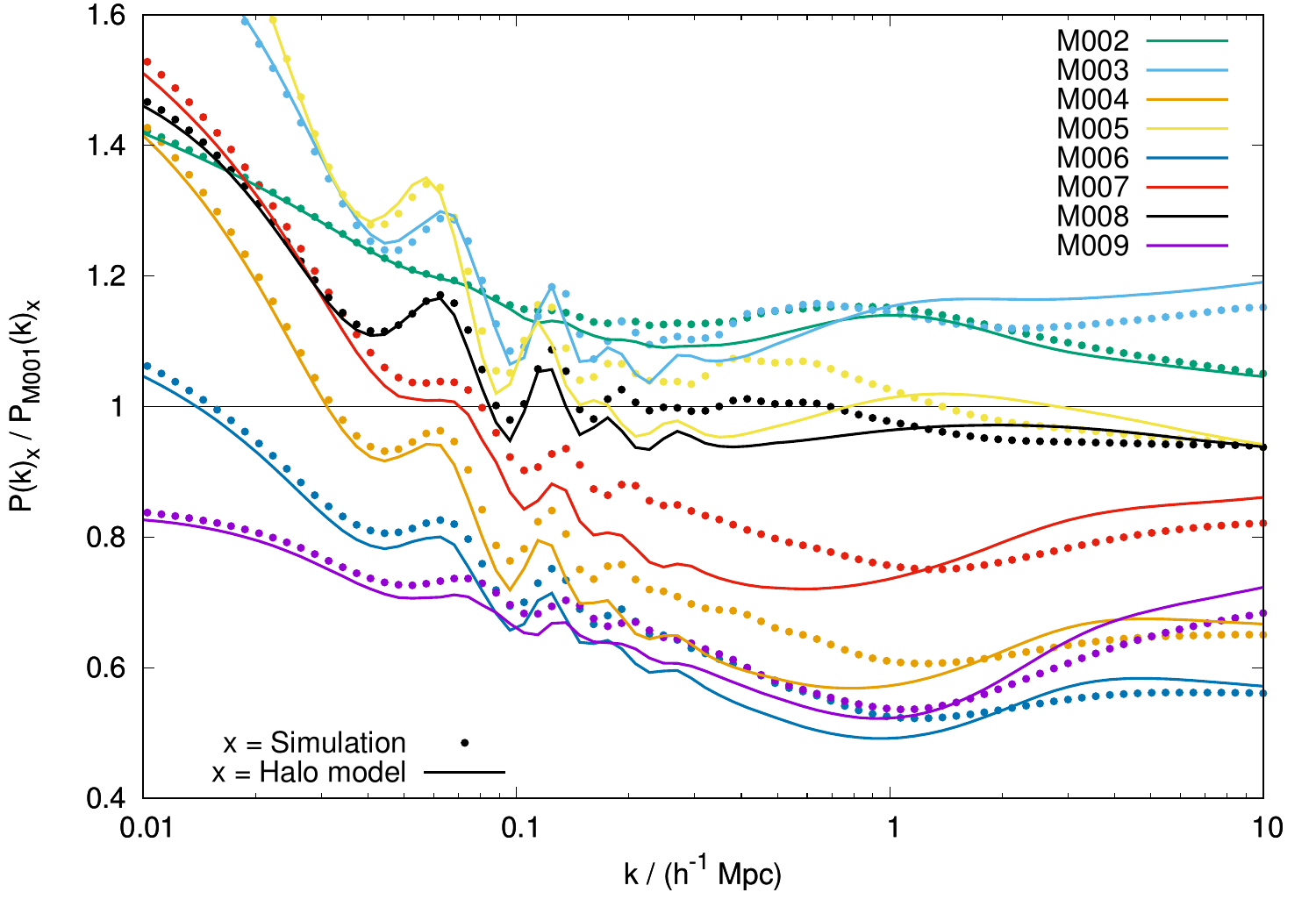}
\end{center}
\caption{The power-spectrum response (the ratio of power to a baseline cosmology; solid lines) of the halo model compared to the same response as measured at $z=0$ in simulations (points) for cosmologies \texttt{M002} to \texttt{M009} of \emu with cosmology \texttt{M001} taken as the baseline. It is well known that the standard halo-model power spectrum provides a poor match to measurements from accurate simulations, but this figure demonstrates that the \emph{response} of the halo model to changes in cosmological parameters is also not accurate when the linear spectra are not identical, with differences at the $\sim 10$ per cent level. The halo-model ingredients are the same as those used to produce the excellent results in Figs.~\ref{fig:power} and \ref{fig:progression}. Note that the worst discrepancies occur around the `quasi-linear scales' ($k=0.1$ to $1\iMpc$), which is the transition between the one- and two-halo terms. At smaller scales the halo model performs better, but is still only accurate to $\sim 5$ per cent. To get improved results it is necessary to work at fixed linear power spectrum.}
\label{fig:scaling}
\end{figure}

The accuracy of the halo-model response is degraded when considering cosmologies with different linear spectrum shapes and amplitudes. In Fig.~\ref{fig:scaling}, I show the power-spectrum response at $z=0$ of the halo model to changes in cosmological parameters compared to the response as measured from simulations of the \emu collaboration (\texttt{M001} to \texttt{M009} from the emulator of \citealt{Heitmann2014}), which span a range of linear spectrum shapes, amplitudes and $w$CDM cosmological parameters. The curves in Fig.~\ref{fig:scaling} are computed as the ratio of a power spectrum to a reference cosmology, which I take as the first `node' (\texttt{M001}) of the \emu simulations. Fig.~\ref{fig:scaling} demonstrates that the way the halo model responds to more general changes in cosmological parameters is inaccurate, with residual differences at the $\sim10$ per cent level. The worst discrepancies occur between $k=0.1$ and $1\iMpc$, in the so-called quasi-linear regime, which suggests that there are additional processes that go into shaping the non-linear spectrum at these scales, the cosmology dependence of which is not captured by the halo model. \new{Indeed, the well-known inaccuracies and inconsistencies of the halo model in the transition regime may not cancel out when one makes these general changes in cosmology. In contrast, from Fig.~\ref{fig:power} one can infer that at fixed linear theory these inaccuracies in the modelling are close to identical and cancel out near perfectly. Taking the example of the transition regime; in the halo model the power here is a sum of the two-halo term which is governed by linear theory and the bias, and the one-halo term which is close to white noise at these scales. At fixed linear theory all of these contributions will be identical (small changes in $\delta_\mathrm{c}$ withstanding), and it follows that the problems of this simplistic modelling will cancel out when the ratio is taken between two cosmologies that share a linear theory.} The greatly improved results seen in Fig.~\ref{fig:power} \new{also} suggest that these quasi-linear processes have an identical effect on the power spectrum if the linear spectrum is identical, which suggests that they must be expressible as functions of the linear spectrum only (or some quantity related to it, such as $\sigma$). This is in accordance with expectations from perturbation theory \citep[\eg][]{Bernardeau2002,Pietroni2008,Taruya2008} where perturbative corrections to the linear spectrum have a very weak dependence on the underlying cosmological parameters for a fixed linear power spectrum. The performance of the halo model is better for $k>1\iMpc$, where discrepancies of only $\sim5$ per cent level are seen, and the gradient of the halo-model prediction is also quite accurate. Fig.~\ref{fig:scaling} was produced using the same halo-model ingredients as Figs.~\ref{fig:power} and~\ref{fig:progression} and of course it is possible that using different ingredients may lead to improved results. In attempts to improve the match in Fig.~\ref{fig:scaling} I investigated different mass functions \citep[\eg][]{Warren2006,Tinker2008} and halo $c(M)$ relations \citep[\eg][]{Duffy2008,Prada2012}, but none lead to any significant improvements.

\section{Summary and discussion}
\label{sec:summary}

It is well known that there are features of the non-linear power spectrum that are not well captured by the standard halo-model calculation, which suffers from a deficit of power at the transition between the one- and two-halo terms, and other inaccuracies at smaller scales. However, in this paper I have demonstrated that incorporating the cosmology dependence of the spherical-collapse model predictions within a halo-model calculation provides an accurate non-linear response in the power spectrum in both real and redshift space. This accuracy is at the per cent level for $k<5\iMpc$ if one restricts oneself to cosmological models that share a linear spectrum in shape and amplitude, but is severely degraded if this restriction is dropped. This suggests that the mechanisms at play in shaping the quasi-linear density field ($k<0.3\iMpc$) should be accurately described as universal functions of the linear power spectrum, which is in agreement with higher-order perturbation theories \citep[\eg][]{Bernardeau2002}. These theories are restricted to the quasi-linear regime (pre shell crossing) and typically have a very weak dependence on the cosmological parameters other than via the linear spectrum. At smaller scales, cosmological hysteresis clearly has an impact on structure growth and must be accounted for in the modelling to provide an accurate description of the deeply non-linear power.

The work presented here indicates that the details of the spherical-collapse calculation are very relevant to the details of structure formation, and that the spherical-collapse model should not simply be treated as a toy that simply provides some rough guidelines. A corollary of this is that identifying haloes according to spherical-collapse predictions may be helpful in generating a truly universal mass function, although it should be remembered that different halo definitions are useful for different tasks. It is possible that reported non-universality \citep[\eg][]{Reed2007,Tinker2008,Courtin2011,Watson2013} could be attributed to identifying haloes in an inconsistent manner. This agrees with the conclusions of \cite{Despali2016}, who report no deviations from mass-function universality for \LCDM if haloes are identified using SO with a cosmology-dependent $\Dv$ and when the cosmology dependence of $\dc$ is included in a fitting function. It would be interesting to see if this reported universality extends to dark-energy models.

This paper demonstrates that the halo model, incorporating the spherical model, may be used as an accurate predictive tool when applied as a correction to simulations. For example, if one is interested in the power spectrum of some strange dark energy cosmology, then all one need do is take an accurate power spectrum for \LCDM with the same spectral shape and amplitude (\eg from \emu) and multiply it by a ratio of halo-model predictions that incorporate the spherical-collapse model. Taking this ratio `cancels' the standard inaccuracies from the halo model that prevent it being used as an accurate tool for the absolute value of the power. Part of the accuracy obtained in this work also stems from using the `correct' concentration--mass relation for haloes; one that takes into account hysteresis such that the halo core retains some memory of the density of the universe at the time when the halo forms. The current prescriptions for the cosmology dependence of the concentration--mass relation are relatively old, and the \cite{Dolag2004} correction for dark energy is relatively crass. Improving these may be a fruitful avenue for further research.

Using the halo model in this way is similar in spirit to methods proposed by \cite{Francis2007} and \cite{Casarini2016}, whereby the power spectrum of $w(a)$CDM models was shown to be accurately matched by a $w$CDM cosmology with identical linear spectrum shape and amplitude and the additional constraint that the conformal times match closely, a scheme that essentially matches the growth history. Cosmologies with similar growth histories will have similar spherical-collapse parameters, and therefore similar non-linear power spectra \citep{Ma2007}, which links to the work in this paper. An advantage of the approach I propose is that it does not require a library of accurate $w$CDM power spectra, and that it may in principle provide accurate spectra for cosmological models other than homogeneous dark energy (\eg coupled dark matter-energy, modified gravity, clustered dark energy). All that is required is an accurate power spectrum for a \LCDM model with the same linear theory, and a spherical-model calculation of $\dc$ and $\Dv$, which is far simpler and less computationally expensive than running full simulations. To reiterate; the accuracy demonstrated in this paper is all achieved with no fitting parameters whatsoever. For the homogeneous dark energy cosmologies discussed in this work the power-spectrum accuracy is one per cent for $k\leq5\iMpc$, and this could conceivably be improved by better concentration--mass modelling, or a more consistent treatment of halo virialization in the spherical model. 

\new{There has been other recent work that attempts to improve the halo-model power calculation directly, and that does not focus only attempt to improve the response. \cite{Mead2015b} add several free parameters to the standard calculation and fit these to high resolution simulations, achieving $5$ per cent accuracy for $k<10\iMpc$. \cite{Mohammed2014a} replace the standard two-halo term with the power spectrum from the \cite{Zeldovich1970} approximation and the standard one-halo term with a power series in $k$. By fitting terms in this power series to \emu data the authors obtain impressive accuracy for $k<0.7\iMpc$. \cite{Seljak2015} pursue a similar method, but use a different expansion for the one-halo term, insist that perturbation theory is matched at low $k$ and that the matter correlation function is simultaneously well predicted. While these latter two methods improve greatly on the basic halo model in terms of power spectrum accuracy I feel that they lose predictive power by throwing away the physical nature of the one-halo term. One of the conclusions of this paper is that changes in power at small scales really do derive from changes to haloes, and these can be accurately predicted via the spherical model; it is unclear how one would integrate this idea with a power-series based one-halo term. However, a fruitful avenue of further research may be to attempt to combine these various recent approaches.}

Obviously this paper does not address the pressing issue of the effect of baryonic feedback on the power spectrum \new{\citep[\eg][]{Semboloni2011,Eifler2015,Fedeli2014a,Mohammed2014b,Harnois-Deraps2015a,Mead2015b}}, but any sensible approach to addressing the impact of feedback will need to work from the basis of an accurate prediction for the power in the absence of baryons. In the future, it would be interesting to see if any of the deeply non-linear features induced in the power spectrum by dark energy can be disentangled from ignorance of feedback effects.

It may be possible to extend the approach presented in this paper to higher order statistics, or other measures of inhomogeneity for which the halo model provides predictions that can be compared to simulations. Examples would be real- and redshift-space correlation functions, bi- or trispectra, weak-lensing peak counts or the power spectra of transformed or clipped density fields.

\section*{Acknowledgements}
I acknowledge support from a CITA National Fellowship and from NSERC. In addition, I am grateful for help from Olivia Steele in modifying \textsc{gadget-2} to support $w(a)$ dark energy models. Useful conversations were had with Catherine Heymans, John Peacock, Alex Hall, Ludovic Van Waerbeke and Sarah Brown. \new{I also acknowledge useful comments and suggestions from an anonymous referee.} The simulations used for this project were enabled in part by support from \href{http://www.westgrid.ca}{WestGrid} and \href{http://www.computecanada.ca}{Compute Canada - Calcul Canada}.

\label{lastpage}

\footnotesize{
\setlength{\bibhang}{2.0em}
\setlength\labelwidth{0.0em}
\bibliographystyle{mnras}
\bibliography{./../meadbib}

\begin{thebibliography}{}
\makeatletter
\relax
\def\mn@urlcharsother{\let\do\@makeother \do\$\do\&\do\#\do\^\do\_\do\%\do\~}
\def\mn@doi{\begingroup\mn@urlcharsother \@ifnextchar [ {\mn@doi@}
  {\mn@doi@[]}}
\def\mn@doi@[#1]#2{\def\@tempa{#1}\ifx\@tempa\@empty \href
  {http://dx.doi.org/#2} {doi:#2}\else \href {http://dx.doi.org/#2} {#1}\fi
  \endgroup}
\def\mn@eprint#1#2{\mn@eprint@#1:#2::\@nil}
\def\mn@eprint@arXiv#1{\href {http://arxiv.org/abs/#1} {{\tt arXiv:#1}}}
\def\mn@eprint@dblp#1{\href {http://dblp.uni-trier.de/rec/bibtex/#1.xml}
  {dblp:#1}}
\def\mn@eprint@#1:#2:#3:#4\@nil{\def\@tempa {#1}\def\@tempb {#2}\def\@tempc
  {#3}\ifx \@tempc \@empty \let \@tempc \@tempb \let \@tempb \@tempa \fi \ifx
  \@tempb \@empty \def\@tempb {arXiv}\fi \@ifundefined
  {mn@eprint@\@tempb}{\@tempb:\@tempc}{\expandafter \expandafter \csname
  mn@eprint@\@tempb\endcsname \expandafter{\@tempc}}}

\bibitem[\protect\citeauthoryear{{Alimi}, {F{\"u}zfa}, {Boucher}, {Rasera},
  {Courtin}  \& {Corasaniti}}{{Alimi} et~al.}{2010}]{Alimi2010}
{Alimi} J.-M.,  {F{\"u}zfa} A.,  {Boucher} V.,  {Rasera} Y.,  {Courtin} J.,
  {Corasaniti} P.-S.,  2010, \mn@doi [\mnras]
  {10.1111/j.1365-2966.2009.15712.x}, \href
  {http://adsabs.harvard.edu/abs/2010MNRAS.401..775A} {401, 775}

\bibitem[\protect\citeauthoryear{{Barreira}, {Li}, {Baugh}  \&
  {Pascoli}}{{Barreira} et~al.}{2013}]{Barreira2013}
{Barreira} A.,  {Li} B.,  {Baugh} C.~M.,   {Pascoli} S.,  2013, \mn@doi [\jcap]
  {10.1088/1475-7516/2013/11/056}, \href
  {http://adsabs.harvard.edu/abs/2013JCAP...11..056B} {11, 56}

\bibitem[\protect\citeauthoryear{Bernardeau, Colombi, Gazta\~{n}aga  \&
  Scoccimarro}{Bernardeau et~al.}{2002}]{Bernardeau2002}
Bernardeau F.,  Colombi S.,  Gazta\~{n}aga E.,   Scoccimarro R.,  2002, \mn@doi
  [Physics Reports] {10.1016/S0370-1573(02)00135-7}, 367, 1

\bibitem[\protect\citeauthoryear{{Bhattacharya}, {Heitmann}, {White},
  {Luki{\'c}}, {Wagner}  \& {Habib}}{{Bhattacharya}
  et~al.}{2011}]{Bhattacharya2011}
{Bhattacharya} S.,  {Heitmann} K.,  {White} M.,  {Luki{\'c}} Z.,  {Wagner} C.,
   {Habib} S.,  2011, \mn@doi [\apj] {10.1088/0004-637X/732/2/122}, \href
  {http://adsabs.harvard.edu/abs/2011ApJ...732..122B} {732, 122}

\bibitem[\protect\citeauthoryear{{Bond}, {Cole}, {Efstathiou}  \&
  {Kaiser}}{{Bond} et~al.}{1991}]{Bond1991}
{Bond} J.~R.,  {Cole} S.,  {Efstathiou} G.,   {Kaiser} N.,  1991, \mn@doi
  [\apj] {10.1086/170520}, \href
  {http://adsabs.harvard.edu/abs/1991ApJ...379..440B} {379, 440}

\bibitem[\protect\citeauthoryear{{Brax}, {Rosenfeld}  \& {Steer}}{{Brax}
  et~al.}{2010}]{Brax2010}
{Brax} P.,  {Rosenfeld} R.,   {Steer} D.~A.,  2010, \mn@doi [\jcap]
  {10.1088/1475-7516/2010/08/033}, \href
  {http://adsabs.harvard.edu/abs/2010JCAP...08..033B} {8, 33}

\bibitem[\protect\citeauthoryear{Bryan \& Norman}{Bryan \&
  Norman}{1998}]{Bryan1998}
Bryan G.~L.,  Norman M.~L.,  1998, \mn@doi [\apj] {10.1086/305262}, 495, 80

\bibitem[\protect\citeauthoryear{Bullock, Kolatt, Sigad, Somerville, Kravtsov,
  Klypin, Primack  \& Dekel}{Bullock et~al.}{2001}]{Bullock2001}
Bullock J.~S.,  Kolatt T.~S.,  Sigad Y.,  Somerville R.~S.,  Kravtsov A.~V.,
  Klypin A.~A.,  Primack J.~R.,   Dekel A.,  2001, \mn@doi [\mnras]
  {10.1046/j.1365-8711.2001.04068.x}, 321, 559

\bibitem[\protect\citeauthoryear{{Casarini}, {Bonometto}, {Tessarotto}  \&
  {Corasaniti}}{{Casarini} et~al.}{2016}]{Casarini2016}
{Casarini} L.,  {Bonometto} S.~A.,  {Tessarotto} E.,   {Corasaniti} P.-S.,
  2016, \mn@doi [\jcap] {10.1088/1475-7516/2016/08/008}, \href
  {http://adsabs.harvard.edu/abs/2016JCAP...08..008C} {8, 008}

\bibitem[\protect\citeauthoryear{{Coles} \& {Lucchin}}{{Coles} \&
  {Lucchin}}{2002}]{b:Coles}
{Coles} P.,  {Lucchin} F.,  2002, {Cosmology: The Origin and Evolution of
  Cosmic Structure, Second Edition}.
Chichester: Wiley

\bibitem[\protect\citeauthoryear{Cooray \& Sheth}{Cooray \&
  Sheth}{2002}]{Cooray2002}
Cooray A.,  Sheth R.,  2002, \mn@doi [Physics Reports]
  {10.1016/S0370-1573(02)00276-4}, 372, 1

\bibitem[\protect\citeauthoryear{{Courtin}, {Rasera}, {Alimi}, {Corasaniti},
  {Boucher}  \& {F{\"u}zfa}}{{Courtin} et~al.}{2011}]{Courtin2011}
{Courtin} J.,  {Rasera} Y.,  {Alimi} J.-M.,  {Corasaniti} P.-S.,  {Boucher} V.,
    {F{\"u}zfa} A.,  2011, \mn@doi [\mnras] {10.1111/j.1365-2966.2010.17573.x},
  \href {http://adsabs.harvard.edu/abs/2011MNRAS.410.1911C} {410, 1911}

\bibitem[\protect\citeauthoryear{{Davis}, {Efstathiou}, {Frenk}  \&
  {White}}{{Davis} et~al.}{1985}]{Davis1985}
{Davis} M.,  {Efstathiou} G.,  {Frenk} C.~S.,   {White} S.~D.~M.,  1985,
  \mn@doi [\apj] {10.1086/163168}, \href
  {http://adsabs.harvard.edu/abs/1985ApJ...292..371D} {292, 371}

\bibitem[\protect\citeauthoryear{{Despali}, {Giocoli}, {Angulo}, {Tormen},
  {Sheth}, {Baso}  \& {Moscardini}}{{Despali} et~al.}{2016}]{Despali2016}
{Despali} G.,  {Giocoli} C.,  {Angulo} R.~E.,  {Tormen} G.,  {Sheth} R.~K.,
  {Baso} G.,   {Moscardini} L.,  2016, \mn@doi [\mnras]
  {10.1093/mnras/stv2842}, \href
  {http://adsabs.harvard.edu/abs/2016MNRAS.456.2486D} {456, 2486}

\bibitem[\protect\citeauthoryear{{Dolag}, {Bartelmann}, {Perrotta},
  {Baccigalupi}, {Moscardini}, {Meneghetti}  \& {Tormen}}{{Dolag}
  et~al.}{2004}]{Dolag2004}
{Dolag} K.,  {Bartelmann} M.,  {Perrotta} F.,  {Baccigalupi} C.,  {Moscardini}
  L.,  {Meneghetti} M.,   {Tormen} G.,  2004, \mn@doi [\aap]
  {10.1051/0004-6361:20031757}, \href
  {http://adsabs.harvard.edu/abs/2004A%26A...416..853D} {416, 853}

\bibitem[\protect\citeauthoryear{{Duffy}, {Schaye}, {Kay}  \& {Dalla
  Vecchia}}{{Duffy} et~al.}{2008}]{Duffy2008}
{Duffy} A.~R.,  {Schaye} J.,  {Kay} S.~T.,   {Dalla Vecchia} C.,  2008, \mn@doi
  [\mnras] {10.1111/j.1745-3933.2008.00537.x}, \href
  {http://adsabs.harvard.edu/abs/2008MNRAS.390L..64D} {390, L64}

\bibitem[\protect\citeauthoryear{{Eifler}, {Krause}, {Dodelson}, {Zentner},
  {Hearin}  \& {Gnedin}}{{Eifler} et~al.}{2015}]{Eifler2015}
{Eifler} T.,  {Krause} E.,  {Dodelson} S.,  {Zentner} A.~R.,  {Hearin} A.~P.,
  {Gnedin} N.~Y.,  2015, \mn@doi [\mnras] {10.1093/mnras/stv2000}, \href
  {http://adsabs.harvard.edu/abs/2015MNRAS.454.2451E} {454, 2451}

\bibitem[\protect\citeauthoryear{{Eke}, {Cole}  \& {Frenk}}{{Eke}
  et~al.}{1996}]{Eke1996}
{Eke} V.~R.,  {Cole} S.,   {Frenk} C.~S.,  1996, \mnras, \href
  {http://adsabs.harvard.edu/abs/1996MNRAS.282..263E} {282, 263}

\bibitem[\protect\citeauthoryear{{Fedeli}}{{Fedeli}}{2014}]{Fedeli2014a}
{Fedeli} C.,  2014, \mn@doi [\jcap] {10.1088/1475-7516/2014/04/028}, \href
  {http://adsabs.harvard.edu/abs/2014JCAP...04..028F} {4, 28}

\bibitem[\protect\citeauthoryear{{Francis}, {Lewis}  \& {Linder}}{{Francis}
  et~al.}{2007}]{Francis2007}
{Francis} M.~J.,  {Lewis} G.~F.,   {Linder} E.~V.,  2007, \mn@doi [\mnras]
  {10.1111/j.1365-2966.2007.12139.x}, \href
  {http://adsabs.harvard.edu/abs/2007MNRAS.380.1079F} {380, 1079}

\bibitem[\protect\citeauthoryear{{Gunn} \& {Gott}}{{Gunn} \&
  {Gott}}{1972}]{Gunn1972}
{Gunn} J.~E.,  {Gott} III J.~R.,  1972, \mn@doi [\apj] {10.1086/151605}, \href
  {http://adsabs.harvard.edu/abs/1972ApJ...176....1G} {176, 1}

\bibitem[\protect\citeauthoryear{{Harnois-D{\'e}raps}, {van Waerbeke}, {Viola}
  \& {Heymans}}{{Harnois-D{\'e}raps} et~al.}{2015}]{Harnois-Deraps2015a}
{Harnois-D{\'e}raps} J.,  {van Waerbeke} L.,  {Viola} M.,   {Heymans} C.,
  2015, \mn@doi [\mnras] {10.1093/mnras/stv646}, \href
  {http://adsabs.harvard.edu/abs/2015MNRAS.450.1212H} {450, 1212}

\bibitem[\protect\citeauthoryear{{Heitmann}, {Lawrence}, {Kwan}, {Habib}  \&
  {Higdon}}{{Heitmann} et~al.}{2014}]{Heitmann2014}
{Heitmann} K.,  {Lawrence} E.,  {Kwan} J.,  {Habib} S.,   {Higdon} D.,  2014,
  \mn@doi [\apj] {10.1088/0004-637X/780/1/111}, \href
  {http://adsabs.harvard.edu/abs/2014ApJ...780..111H} {780, 111}

\bibitem[\protect\citeauthoryear{{Ichiki} \& {Takada}}{{Ichiki} \&
  {Takada}}{2012}]{Ichiki2012}
{Ichiki} K.,  {Takada} M.,  2012, \mn@doi [\prd] {10.1103/PhysRevD.85.063521},
  \href {http://adsabs.harvard.edu/abs/2012PhRvD..85f3521I} {85, 063521}

\bibitem[\protect\citeauthoryear{Jenkins, Frenk, White, Colberg, Cole, Evrard,
  Couchman  \& Yoshida}{Jenkins et~al.}{2001}]{Jenkins2001}
Jenkins A.,  Frenk C.~S.,  White S. D.~M.,  Colberg J.~M.,  Cole S.,  Evrard
  A.~E.,  Couchman H. M.~P.,   Yoshida N.,  2001, \mn@doi [\mnras]
  {10.1046/j.1365-8711.2001.04029.x}, 321, 372

\bibitem[\protect\citeauthoryear{{Kaiser}}{{Kaiser}}{1987}]{Kaiser1987}
{Kaiser} N.,  1987, \mnras, \href
  {http://adsabs.harvard.edu/abs/1987MNRAS.227....1K} {227, 1}

\bibitem[\protect\citeauthoryear{{Kazantzidis}, {Magorrian}  \&
  {Moore}}{{Kazantzidis} et~al.}{2004}]{Kazantzidis2004}
{Kazantzidis} S.,  {Magorrian} J.,   {Moore} B.,  2004, \mn@doi [\apj]
  {10.1086/380192}, \href {http://adsabs.harvard.edu/abs/2004ApJ...601...37K}
  {601, 37}

\bibitem[\protect\citeauthoryear{{Knebe} et~al.,}{{Knebe}
  et~al.}{2011}]{Knebe2011}
{Knebe} A.,  et~al., 2011, \mn@doi [\mnras] {10.1111/j.1365-2966.2011.18858.x},
  \href {http://adsabs.harvard.edu/abs/2011MNRAS.415.2293K} {415, 2293}

\bibitem[\protect\citeauthoryear{{Lacey} \& {Cole}}{{Lacey} \&
  {Cole}}{1993}]{Lacey1993}
{Lacey} C.,  {Cole} S.,  1993, \mnras, \href
  {http://adsabs.harvard.edu/abs/1993MNRAS.262..627L} {262, 627}

\bibitem[\protect\citeauthoryear{{Lahav}, {Lilje}, {Primack}  \&
  {Rees}}{{Lahav} et~al.}{1991}]{Lahav1991}
{Lahav} O.,  {Lilje} P.~B.,  {Primack} J.~R.,   {Rees} M.~J.,  1991, \mn@doi
  [\mnras] {10.1093/mnras/251.1.128}, \href
  {http://adsabs.harvard.edu/abs/1991MNRAS.251..128L} {251, 128}

\bibitem[\protect\citeauthoryear{Lewis, Challinor  \& Lasenby}{Lewis
  et~al.}{2000}]{Lewis2000}
Lewis A.,  Challinor A.,   Lasenby A.,  2000, \mn@doi [\apj] {10.1086/309179},
  538, 473

\bibitem[\protect\citeauthoryear{{Li} \& {Efstathiou}}{{Li} \&
  {Efstathiou}}{2012}]{Li2012b}
{Li} B.,  {Efstathiou} G.,  2012, \mn@doi [\mnras]
  {10.1111/j.1365-2966.2011.20404.x}, \href
  {http://adsabs.harvard.edu/abs/2012MNRAS.421.1431L} {421, 1431}

\bibitem[\protect\citeauthoryear{{LoVerde}}{{LoVerde}}{2014}]{LoVerde2014}
{LoVerde} M.,  2014, \mn@doi [\prd] {10.1103/PhysRevD.90.083518}, \href
  {http://adsabs.harvard.edu/abs/2014PhRvD..90h3518L} {90, 083518}

\bibitem[\protect\citeauthoryear{{Lynden-Bell}}{{Lynden-Bell}}{1967}]{Lynden-Bell1967}
{Lynden-Bell} D.,  1967, \mnras, \href
  {http://adsabs.harvard.edu/abs/1967MNRAS.136..101L} {136, 101}

\bibitem[\protect\citeauthoryear{{Ma}}{{Ma}}{2007}]{Ma2007}
{Ma} Z.,  2007, \mn@doi [\apj] {10.1086/519440}, \href
  {http://adsabs.harvard.edu/abs/2007ApJ...665..887M} {665, 887}

\bibitem[\protect\citeauthoryear{{Maor} \& {Lahav}}{{Maor} \&
  {Lahav}}{2005}]{Maor2005}
{Maor} I.,  {Lahav} O.,  2005, \mn@doi [\jcap] {10.1088/1475-7516/2005/07/003},
  \href {http://adsabs.harvard.edu/abs/2005JCAP...07..003M} {7, 003}

\bibitem[\protect\citeauthoryear{{McDonald}, {Trac}  \& {Contaldi}}{{McDonald}
  et~al.}{2006}]{McDonald2006}
{McDonald} P.,  {Trac} H.,   {Contaldi} C.,  2006, \mn@doi [\mnras]
  {10.1111/j.1365-2966.2005.09881.x}, \href
  {http://adsabs.harvard.edu/abs/2006MNRAS.366..547M} {366, 547}

\bibitem[\protect\citeauthoryear{{Mead}, {Peacock}, {Heymans}, {Joudaki}  \&
  {Heavens}}{{Mead} et~al.}{2015}]{Mead2015b}
{Mead} A.~J.,  {Peacock} J.~A.,  {Heymans} C.,  {Joudaki} S.,   {Heavens}
  A.~F.,  2015, \mn@doi [\mnras] {10.1093/mnras/stv2036}, \href
  {http://adsabs.harvard.edu/abs/2015MNRAS.454.1958M} {454, 1958}

\bibitem[\protect\citeauthoryear{{Mead}, {Heymans}, {Lombriser}, {Peacock},
  {Steele}  \& {Winther}}{{Mead} et~al.}{2016}]{Mead2016}
{Mead} A.~J.,  {Heymans} C.,  {Lombriser} L.,  {Peacock} J.~A.,  {Steele}
  O.~I.,   {Winther} H.~A.,  2016, \mn@doi [\mnras] {10.1093/mnras/stw681},
  \href {http://adsabs.harvard.edu/abs/2016MNRAS.459.1468M} {459, 1468}

\bibitem[\protect\citeauthoryear{{Mohammed} \& {Seljak}}{{Mohammed} \&
  {Seljak}}{2014}]{Mohammed2014a}
{Mohammed} I.,  {Seljak} U.,  2014, \mn@doi [\mnras] {10.1093/mnras/stu1972},
  \href {http://adsabs.harvard.edu/abs/2014MNRAS.445.3382M} {445, 3382}

\bibitem[\protect\citeauthoryear{{Mohammed}, {Martizzi}, {Teyssier}  \&
  {Amara}}{{Mohammed} et~al.}{2014}]{Mohammed2014b}
{Mohammed} I.,  {Martizzi} D.,  {Teyssier} R.,   {Amara} A.,  2014, preprint
  (arXiv:e-prints 1410.6826), \href
  {http://adsabs.harvard.edu/abs/2014arXiv1410.6826M} {}

\bibitem[\protect\citeauthoryear{{More}, {Kravtsov}, {Dalal}  \&
  {Gottl{\"o}ber}}{{More} et~al.}{2011}]{More2011}
{More} S.,  {Kravtsov} A.~V.,  {Dalal} N.,   {Gottl{\"o}ber} S.,  2011, \mn@doi
  [ApJS] {10.1088/0067-0049/195/1/4}, \href
  {http://adsabs.harvard.edu/abs/2011ApJS..195....4M} {195, 4}

\bibitem[\protect\citeauthoryear{{Mota} \& {van de Bruck}}{{Mota} \& {van de
  Bruck}}{2004}]{Mota2004}
{Mota} D.~F.,  {van de Bruck} C.,  2004, \mn@doi [\aap]
  {10.1051/0004-6361:20041090}, \href
  {http://adsabs.harvard.edu/abs/2004A%26A...421...71M} {421, 71}

\bibitem[\protect\citeauthoryear{{Nakamura} \& {Suto}}{{Nakamura} \&
  {Suto}}{1997}]{Nakamura1997}
{Nakamura} T.~T.,  {Suto} Y.,  1997, \mn@doi [Progress of Theoretical Physics]
  {10.1143/PTP.97.49}, \href
  {http://adsabs.harvard.edu/abs/1997PThPh..97...49N} {97, 49}

\bibitem[\protect\citeauthoryear{Navarro, Frenk  \& White}{Navarro
  et~al.}{1997}]{Navarro1997}
Navarro J.~F.,  Frenk C.~S.,   White S. D.~M.,  1997, \mn@doi [\apj]
  {10.1086/304888}, 490, 493

\bibitem[\protect\citeauthoryear{{Neto} et~al.,}{{Neto}
  et~al.}{2007}]{Neto2007}
{Neto} A.~F.,  et~al., 2007, \mn@doi [\mnras]
  {10.1111/j.1365-2966.2007.12381.x}, \href
  {http://adsabs.harvard.edu/abs/2007MNRAS.381.1450N} {381, 1450}

\bibitem[\protect\citeauthoryear{{Nusser} \& {Colberg}}{{Nusser} \&
  {Colberg}}{1998}]{Nusser1998}
{Nusser} A.,  {Colberg} J.~M.,  1998, \mn@doi [\mnras]
  {10.1046/j.1365-8711.1998.01218.x}, \href
  {http://adsabs.harvard.edu/abs/1998MNRAS.294..457N} {294, 457}

\bibitem[\protect\citeauthoryear{{Pace}, {Waizmann}  \& {Bartelmann}}{{Pace}
  et~al.}{2010}]{Pace2010}
{Pace} F.,  {Waizmann} J.-C.,   {Bartelmann} M.,  2010, \mn@doi [\mnras]
  {10.1111/j.1365-2966.2010.16841.x}, \href
  {http://adsabs.harvard.edu/abs/2010MNRAS.406.1865P} {406, 1865}

\bibitem[\protect\citeauthoryear{{Peacock} \& {Smith}}{{Peacock} \&
  {Smith}}{2000}]{Peacock2000}
{Peacock} J.~A.,  {Smith} R.~E.,  2000, \mn@doi [\mnras]
  {10.1046/j.1365-8711.2000.03779.x}, \href
  {http://adsabs.harvard.edu/abs/2000MNRAS.318.1144P} {318, 1144}

\bibitem[\protect\citeauthoryear{{Percival}}{{Percival}}{2005}]{Percival2005}
{Percival} W.~J.,  2005, \mn@doi [\aap] {10.1051/0004-6361:20053637}, \href
  {http://adsabs.harvard.edu/abs/2005A%26A...443..819P} {443, 819}

\bibitem[\protect\citeauthoryear{{Pettorino} \& {Baccigalupi}}{{Pettorino} \&
  {Baccigalupi}}{2008}]{Pettorino2008}
{Pettorino} V.,  {Baccigalupi} C.,  2008, \mn@doi [\prd]
  {10.1103/PhysRevD.77.103003}, \href
  {http://adsabs.harvard.edu/abs/2008PhRvD..77j3003P} {77, 103003}

\bibitem[\protect\citeauthoryear{{Pietroni}}{{Pietroni}}{2008}]{Pietroni2008}
{Pietroni} M.,  2008, \mn@doi [\jcap] {10.1088/1475-7516/2008/10/036}, \href
  {http://adsabs.harvard.edu/abs/2008JCAP...10..036P} {10, 036}

\bibitem[\protect\citeauthoryear{{Planck Collaboration XIV}}{{Planck
  Collaboration XIV}}{2016}]{Planck2015XIV}
{Planck Collaboration XIV} 2016, \mn@doi [\aap] {10.1051/0004-6361/201525814},
  \href {http://adsabs.harvard.edu/abs/2016A%26A...594A..14P} {594, A14}

\bibitem[\protect\citeauthoryear{{Prada}, {Klypin}, {Cuesta}, {Betancort-Rijo}
  \& {Primack}}{{Prada} et~al.}{2012}]{Prada2012}
{Prada} F.,  {Klypin} A.~A.,  {Cuesta} A.~J.,  {Betancort-Rijo} J.~E.,
  {Primack} J.,  2012, \mn@doi [\mnras] {10.1111/j.1365-2966.2012.21007.x},
  \href {http://adsabs.harvard.edu/abs/2012MNRAS.423.3018P} {423, 3018}

\bibitem[\protect\citeauthoryear{Press \& Schechter}{Press \&
  Schechter}{1974}]{Press1974}
Press W.~H.,  Schechter P.,  1974, \mn@doi [\apj] {10.1086/152650}, 187, 425

\bibitem[\protect\citeauthoryear{{Reed}, {Bower}, {Frenk}, {Jenkins}  \&
  {Theuns}}{{Reed} et~al.}{2007}]{Reed2007}
{Reed} D.~S.,  {Bower} R.,  {Frenk} C.~S.,  {Jenkins} A.,   {Theuns} T.,  2007,
  \mn@doi [\mnras] {10.1111/j.1365-2966.2006.11204.x}, \href
  {http://adsabs.harvard.edu/abs/2007MNRAS.374....2R} {374, 2}

\bibitem[\protect\citeauthoryear{{Schmidt}, {Lima}, {Oyaizu}  \&
  {Hu}}{{Schmidt} et~al.}{2009}]{Schmidt2009a}
{Schmidt} F.,  {Lima} M.,  {Oyaizu} H.,   {Hu} W.,  2009, \mn@doi [\prd]
  {10.1103/PhysRevD.79.083518}, \href
  {http://adsabs.harvard.edu/abs/2009PhRvD..79h3518S} {79, 083518}

\bibitem[\protect\citeauthoryear{{Schmidt}, {Hu}  \& {Lima}}{{Schmidt}
  et~al.}{2010}]{Schmidt2010a}
{Schmidt} F.,  {Hu} W.,   {Lima} M.,  2010, \mn@doi [\prd]
  {10.1103/PhysRevD.81.063005}, \href
  {http://adsabs.harvard.edu/abs/2010PhRvD..81f3005S} {81, 063005}

\bibitem[\protect\citeauthoryear{{Seljak}}{{Seljak}}{2000}]{Seljak2000}
{Seljak} U.,  2000, \mn@doi [\mnras] {10.1046/j.1365-8711.2000.03715.x}, \href
  {http://adsabs.harvard.edu/abs/2000MNRAS.318..203S} {318, 203}

\bibitem[\protect\citeauthoryear{{Seljak} \& {Vlah}}{{Seljak} \&
  {Vlah}}{2015}]{Seljak2015}
{Seljak} U.,  {Vlah} Z.,  2015, \mn@doi [\prd] {10.1103/PhysRevD.91.123516},
  \href {http://adsabs.harvard.edu/abs/2015PhRvD..91l3516S} {91, 123516}

\bibitem[\protect\citeauthoryear{{Semboloni}, {Hoekstra}, {Schaye}, {van
  Daalen}  \& {McCarthy}}{{Semboloni} et~al.}{2011}]{Semboloni2011}
{Semboloni} E.,  {Hoekstra} H.,  {Schaye} J.,  {van Daalen} M.~P.,   {McCarthy}
  I.~G.,  2011, \mn@doi [\mnras] {10.1111/j.1365-2966.2011.19385.x}, \href
  {http://adsabs.harvard.edu/abs/2011MNRAS.417.2020S} {417, 2020}

\bibitem[\protect\citeauthoryear{Sheth \& Tormen}{Sheth \&
  Tormen}{1999}]{Sheth1999}
Sheth R.~K.,  Tormen G.,  1999, \mn@doi [\mnras]
  {10.1046/j.1365-8711.1999.02692.x}, 308, 119

\bibitem[\protect\citeauthoryear{Sheth, Mo  \& Tormen}{Sheth
  et~al.}{2001}]{Sheth2001}
Sheth R.~K.,  Mo H.~J.,   Tormen G.,  2001, \mn@doi [\mnras]
  {10.1046/j.1365-8711.2001.04006.x}, 323, 1

\bibitem[\protect\citeauthoryear{{Smith} et~al.,}{{Smith}
  et~al.}{2003}]{Smith2003}
{Smith} R.~E.,  et~al., 2003, \mn@doi [\mnras]
  {10.1046/j.1365-8711.2003.06503.x}, \href
  {http://adsabs.harvard.edu/abs/2003MNRAS.341.1311S} {341, 1311}

\bibitem[\protect\citeauthoryear{{Springel}}{{Springel}}{2005}]{Springel2005b}
{Springel} V.,  2005, \mn@doi [\mnras] {10.1111/j.1365-2966.2005.09655.x},
  \href {http://adsabs.harvard.edu/abs/2005MNRAS.364.1105S} {364, 1105}

\bibitem[\protect\citeauthoryear{{Takahashi}, {Sato}, {Nishimichi}, {Taruya}
  \& {Oguri}}{{Takahashi} et~al.}{2012}]{Takahashi2012}
{Takahashi} R.,  {Sato} M.,  {Nishimichi} T.,  {Taruya} A.,   {Oguri} M.,
  2012, \mn@doi [\apj] {10.1088/0004-637X/761/2/152}, \href
  {http://adsabs.harvard.edu/abs/2012ApJ...761..152T} {761, 152}

\bibitem[\protect\citeauthoryear{{Tarrant}, {van de Bruck}, {Copeland}  \&
  {Green}}{{Tarrant} et~al.}{2012}]{Tarrant2012}
{Tarrant} E.~R.~M.,  {van de Bruck} C.,  {Copeland} E.~J.,   {Green} A.~M.,
  2012, \mn@doi [\prd] {10.1103/PhysRevD.85.023503}, \href
  {http://adsabs.harvard.edu/abs/2012PhRvD..85b3503T} {85, 023503}

\bibitem[\protect\citeauthoryear{{Taruya} \& {Hiramatsu}}{{Taruya} \&
  {Hiramatsu}}{2008}]{Taruya2008}
{Taruya} A.,  {Hiramatsu} T.,  2008, \mn@doi [\apj] {10.1086/526515}, \href
  {http://adsabs.harvard.edu/abs/2008ApJ...674..617T} {674, 617}

\bibitem[\protect\citeauthoryear{{Tinker}, {Kravtsov}, {Klypin}, {Abazajian},
  {Warren}, {Yepes}, {Gottl{\"o}ber}  \& {Holz}}{{Tinker}
  et~al.}{2008}]{Tinker2008}
{Tinker} J.,  {Kravtsov} A.~V.,  {Klypin} A.,  {Abazajian} K.,  {Warren} M.,
  {Yepes} G.,  {Gottl{\"o}ber} S.,   {Holz} D.~E.,  2008, \mn@doi [\apj]
  {10.1086/591439}, \href {http://adsabs.harvard.edu/abs/2008ApJ...688..709T}
  {688, 709}

\bibitem[\protect\citeauthoryear{{Valageas} \& {Nishimichi}}{{Valageas} \&
  {Nishimichi}}{2011}]{Valageas2011}
{Valageas} P.,  {Nishimichi} T.,  2011, \mn@doi [\aap]
  {10.1051/0004-6361/201015685}, \href
  {http://adsabs.harvard.edu/abs/2011A%26A...527A..87V} {527, A87}

\bibitem[\protect\citeauthoryear{{Warren}, {Abazajian}, {Holz}  \&
  {Teodoro}}{{Warren} et~al.}{2006}]{Warren2006}
{Warren} M.~S.,  {Abazajian} K.,  {Holz} D.~E.,   {Teodoro} L.,  2006, \mn@doi
  [APJ] {10.1086/504962}, \href
  {http://adsabs.harvard.edu/abs/2006ApJ...646..881W} {646, 881}

\bibitem[\protect\citeauthoryear{{Watson}, {Iliev}, {D'Aloisio}, {Knebe},
  {Shapiro}  \& {Yepes}}{{Watson} et~al.}{2013}]{Watson2013}
{Watson} W.~A.,  {Iliev} I.~T.,  {D'Aloisio} A.,  {Knebe} A.,  {Shapiro} P.~R.,
    {Yepes} G.,  2013, \mn@doi [\mnras] {10.1093/mnras/stt791}, \href
  {http://adsabs.harvard.edu/abs/2013MNRAS.433.1230W} {433, 1230}

\bibitem[\protect\citeauthoryear{{Weller} \& {Lewis}}{{Weller} \&
  {Lewis}}{2003}]{Weller2003}
{Weller} J.,  {Lewis} A.~M.,  2003, \mn@doi [\mnras]
  {10.1111/j.1365-2966.2003.07144.x}, \href
  {http://adsabs.harvard.edu/abs/2003MNRAS.346..987W} {346, 987}

\bibitem[\protect\citeauthoryear{{White}}{{White}}{2001}]{White2001}
{White} M.,  2001, \mn@doi [\mnras] {10.1046/j.1365-8711.2001.03956.x}, \href
  {http://adsabs.harvard.edu/abs/2001MNRAS.321....1W} {321, 1}

\bibitem[\protect\citeauthoryear{{Wintergerst} \& {Pettorino}}{{Wintergerst} \&
  {Pettorino}}{2010}]{Wintergerst2010}
{Wintergerst} N.,  {Pettorino} V.,  2010, \mn@doi [\prd]
  {10.1103/PhysRevD.82.103516}, \href
  {http://adsabs.harvard.edu/abs/2010PhRvD..82j3516W} {82, 103516}

\bibitem[\protect\citeauthoryear{{Zel'dovich}}{{Zel'dovich}}{1970}]{Zeldovich1970}
{Zel'dovich} Y.~B.,  1970, AAP, \href
  {http://adsabs.harvard.edu/abs/1970A%26A.....5...84Z} {5, 84}

\bibitem[\protect\citeauthoryear{{Zheng}, {Tinker}, {Weinberg}  \&
  {Berlind}}{{Zheng} et~al.}{2002}]{Zheng2002}
{Zheng} Z.,  {Tinker} J.~L.,  {Weinberg} D.~H.,   {Berlind} A.~A.,  2002,
  \mn@doi [\apj] {10.1086/341434}, \href
  {http://adsabs.harvard.edu/abs/2002ApJ...575..617Z} {575, 617}

\makeatother
\end{thebibliography}
}

\normalsize
\appendix

\section{Fitting functions for $\dc$ and $\Dv$}
\label{app:fitting_functions}

\begin{figure*}
\begin{center}
\includegraphics[angle=270,width=18cm]{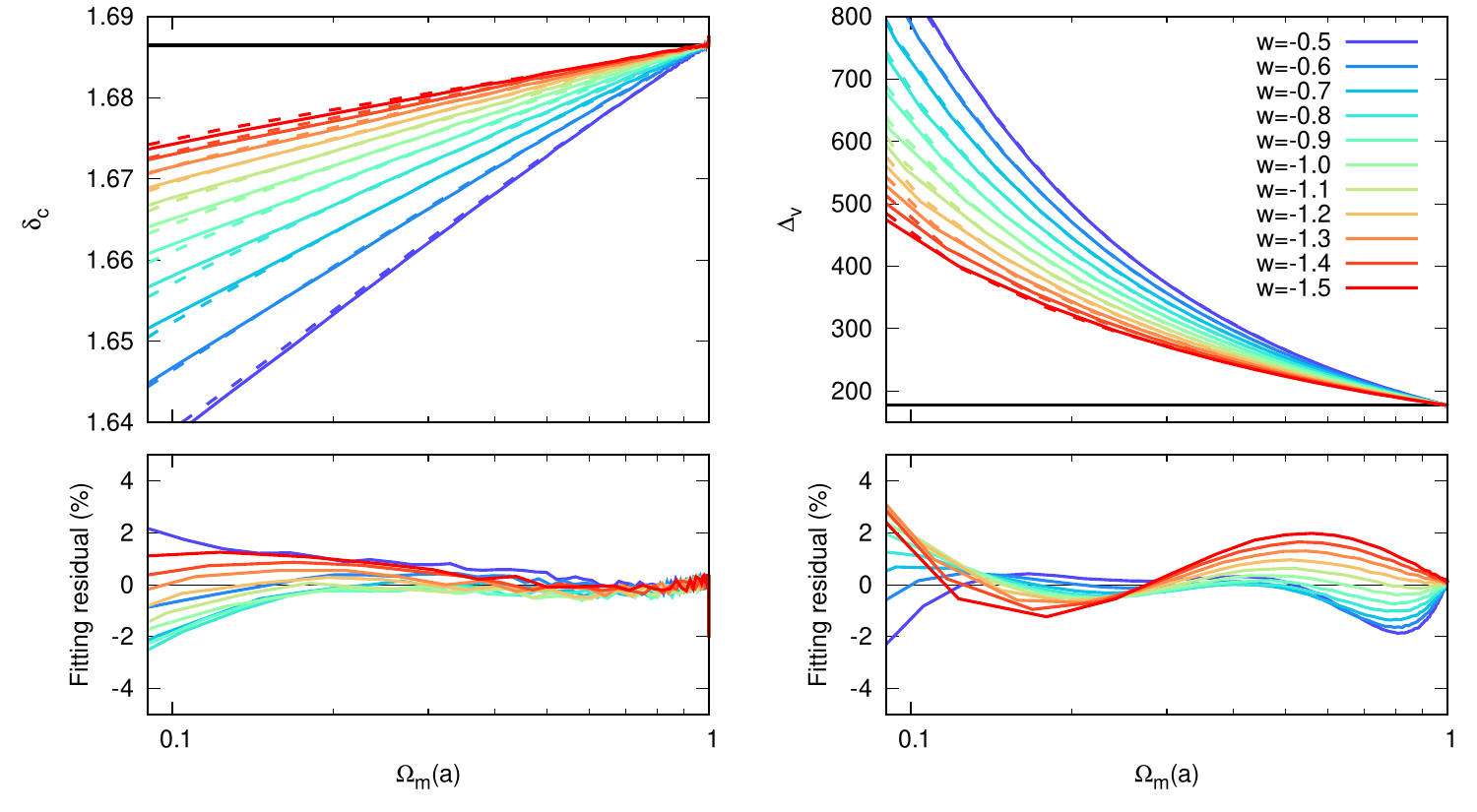}
\end{center}
\caption{The variation of $\dc$ and $\Dv$ with $\Om$ for $w$CDM models; solid lines in the top panels show results from a numerical calculation and the dashed lines show the fitting functions in equations~(\ref{eq:dc_fit}) and (\ref{eq:Dv_fit}). The lower panels show the residuals of the fitting function for each model, but rescaled over the relevant range: $1.64\rightarrow1.69$ for $\dc$ and $150\rightarrow 800$ for $\Dv$. Failing to do this would make the residual of the $\dc$ fitting formula sound misleadingly good because the absolute value changes very little with cosmological model. The parameters of the fitting functions were fitted over a wide range of $w(a)$CDM cosmologies, not only the models shown in this figure.}
\label{fig:dcDv_fit}
\end{figure*}

In this Appendix I provide some fitting functions for the spherical-collapse parameters $\dc$ and $\Dv$ that are valid for homogeneous dark energy models and that even work for substantial curvature. A \LCDM fitting function for $\dc$ can be found in \cite{Nakamura1997} and for $\Dv$ in \cite{Bryan1998}. For matter-dominated open and closed cosmologies exact formulae may be derived \citep[\eg][]{b:Coles}. The fitting functions presented here are constructed in such a way that they do not depend explicitly on the dark-energy parameterization, and are instead parameterized in terms of the matter density, linear growth and the integrated growth, which are defined below.

I parameterize a fitting function for $\dc$ via
\begin{equation}
\frac{\dc}{1.686}=1+\sum_{i=1,2} f_{i}\left(\frac{g}{a},\frac{G}{a}\right)\left[\log_{10}\Om(a)\right]^{\alpha_i}\ ,
\label{eq:dc_fit}
\end{equation}
and for $\Dv$ via
\begin{equation}
\frac{\Dv}{178}=1+\sum_{i=3,4} f_{i}\left(\frac{g}{a},\frac{G}{a}\right)\left[\log_{10}\Om(a)\right]^{\alpha_i}\ ,
\label{eq:Dv_fit}
\end{equation}
with $f_i$ being the quadratics 
\begin{equation}
f_i (x,y)=p_{i,0}+p_{i,1}(1-x)+p_{i,2}(1-x)^2+p_{i,3}(1-y)\ .
\label{eq:quadratic}
\end{equation}
These functions ensure that the Einstein-de Sitter values $\dc=1.686$ and $\Dv=178$ are recovered when $\Om(a)=1$. $g(a)$ is the \emph{unnormalized} linear-theory growth function; the solution to
\begin{equation}
g''+\left(\frac{A}{H^2}+2\right)\frac{g'}{a}=\frac{3}{2}\Om(a)\frac{g}{a^2}\ ,
\label{eq:growth}
\end{equation}
where the dashes represent derivatives with respect to $a$, and taking the initial condition $g=a$ when $a\ll 1$, which is the Einstein-de Sitter growing mode solution that is valid at early times for all cosmologies considered here. $A\equiv\ddot{a}/a$ and $H$ is the standard Hubble parameter. I also define the integrated growth via
\begin{equation}
G(a)=\int_0^a\frac{g(\tilde a)}{\tilde a}\,\mathrm{d}\tilde a\ ,
\label{eq:integrated_growth}
\end{equation}
where the tilde denotes dummy variables. For the Einstein-de Sitter growing mode $G(a)=a$.

\begin{table}
\begin{center}
\caption{Best-fitting parameters for the $\dc$ and $\Dv$ fitting functions in equations~(\ref{eq:dc_fit}) and (\ref{eq:Dv_fit}). The $\alpha_i$ exponents are defined in these equations while the $p_{i,j}$ coefficients of the quadratics are defined in equation~(\ref{eq:quadratic}).}
\begin{tabular}{c c c c}
\hline
$\dc$ & Value & $\Dv$ & Value\\
\hline
$p_{1,0}$ & $-0.0069$ & $p_{3,0}$ & $-0.79$ \\
$p_{1,1}$ & $-0.0208$ & $p_{3,1}$ & $-10.17$\\
$p_{1,2}$ & $0.0312$ & $p_{3,2}$ & $2.51$\\
$p_{1,3}$ & $0.0021$ & $p_{3,3}$ & $6.51$\\
$\alpha_1$ & $1$ & $\alpha_3$ & $1$\\
$p_{2,0}$ & $0.0001$ & $p_{4,0}$ & $-1.89$\\
$p_{2,1}$ & $-0.0647$ & $p_{4,1}$ & $0.38$\\
$p_{2,2}$ & $-0.0417$ & $p_{4,2}$ & $18.8$\\
$p_{2,3}$ & $0.0646$ & $p_{4,3}$ & $-15.87$\\
$\alpha_2$  & $0$ & $\alpha_4$ & $2$ \\
\hline
\end{tabular}
\label{tab:fit_params}
\end{center}
\end{table}

Parameterizing the fitting function in terms of $g(a)/a$ works quite well at capturing the cosmology dependence as the universe deviates from Einstein-de Sitter form, and $G(a)/a$ is necessary to capture the hysteresis. For $\Om=0.3$ \LCDM $g(a=1)\simeq0.779$ and $G(a=1)\simeq0.930$. The best-fitting exponents $\alpha_i$, and coefficients of the quadratics $f_i$, are given in Table~\ref{tab:fit_params}, note that there are a total of $8$ fitted parameters for each of $\dc$ and $\Dv$ and that the $\alpha$ parameters were fixed with some guess work. The quality of the fitting functions for $w$CDM models can be seen in Fig.~\ref{fig:dcDv_fit}. In the top two panels the solid curves show results from a numerical calculation while the dashed lines show the fitting function. In the lower panels, I show the residual, which is a maximum of two per cent for all models shown for $\Om>0.2$. This residual is calculated over the limited range of $\dc$ and $\Dv$ on the axes of the top two panels, rather than being an absolute residual. Quoting the absolute value for the residual would make the $\dc$ fitting function sound anomalously good because the range over which $\dc$ changes is very limited ($\sim1.64\to1.69$) compared to the absolute value. The quality of the fitting function is similar for $w(a)$CDM models and even works for substantial curvature, but these extra models are not shown to avoid confusing the plot. 

If these fitting functions are not of sufficient accuracy then the code I wrote for the spherical-collapse calculation used in this work is available at \meadaddress.

\end{document}